\documentclass[11pt]{article}
 \pdfoutput = 1
 
 \usepackage[utf8]{inputenc}
 \usepackage{color,graphicx}
 \usepackage{epsfig}
 \usepackage{amsmath}
 \usepackage{verbatim}
 \usepackage{amssymb}
 \usepackage{mathabx}
 \usepackage[mathcal]{eucal}
 \usepackage{stmaryrd}
 \usepackage{empheq}
 \usepackage{esint}
 \usepackage{physics}
 \usepackage{xcolor}
 \usepackage{amsfonts}
 \usepackage{cite}
 \usepackage{array}
 \usepackage{setspace}
 \usepackage{braket}
 \usepackage{float}
 \usepackage{url}
 \usepackage{mathtools}
 \usepackage{tensor}
 \usepackage{bbold}
 \usepackage{slashed}
 \usepackage{tikz}
 \usetikzlibrary{decorations}
 \usepackage{graphicx}
 \usepackage{epstopdf}
 \usepackage{subcaption}
 \usepackage[labelfont=bf,font={sf}]{caption}
 \usepackage{pgfplots}
 \usepackage{mathrsfs}
 \usepackage{fancybox}
 \usepackage{eurosym}
 \usepackage{tcolorbox}
 \usepackage{tensor}
 \usepackage[normalem]{ulem}
 \allowdisplaybreaks
 \usepackage{changepage}

 \usepackage[margin = 2.2cm]{geometry}
 \usepackage[ragged]{footmisc}
 \usepackage[bookmarks=true,bookmarksnumbered=false,
 hyperindex=true,bookmarksopen=true,hyperfigures=true,
 colorlinks=true,linkcolor=webblue,citecolor=webgreen,urlcolor=webblue,breaklinks]{hyperref}
 \definecolor{webgreen}{rgb}{0, 0.5, 0}
 \definecolor{webblue}{rgb}{0, 0, 0.5}
 \definecolor{webred}{rgb}{0.5, 0, 0}
 \definecolor{darkgreen}{rgb}{0,0.5,0}

 \definecolor{mygenta}{RGB}{255,0,255}

 \def\ben{\begin{equation}}
 \def\een{\end{equation}}

    \let\d=\delta 
    
      \let\r=v

 \def\be{\begin{equation}}
 \def\ee{\end{equation}}
 \def\ba{\begin{array}}
 \def\ea{\end{array}}

 \def\dalemb#1#2{{\vbox{\hrule height .#2pt
 \hbox{\vrule width.#2pt height#1pt \kern#1pt
 \vrule width.#2pt}
 \hrule height.#2pt}}}
 
 \newcommand{\bea}{\begin{eqnarray}}
 \newcommand{\eea}{\end{eqnarray}}
 \def\eps{{\epsilon}}

 \renewcommand{\d}{\mathrm{d}}
 \renewcommand{\i}{\mathrm{i}}

 \definecolor{Blue}{HTML}{0072B2}
\definecolor{Orange}{HTML}{E69F00}
\definecolor{Green}{HTML}{66B300}

 \numberwithin{equation}{section}

\begin{document}
 
\thispagestyle{empty}
 ~\vspace{5mm}
\begin{adjustwidth}{-1cm}{-1cm}
\begin{center}
 {\LARGE \bf 
The cosmological necklace problem
}
 \vspace{0.4in}

 {\bf Andreas Blommaert${}^{1,2}$, Jonah Kudler-Flam${}^{1}$, Vladimir Narovlansky${}^{3,4}$, and Erez Y. Urbach${}^{1,5}$}
 \end{center}
 \end{adjustwidth}
\begin{center}
 \vspace{0.4in}
 {${}^1$School of Natural Sciences, Institute for Advanced Study, Princeton, NJ 08540, USA}

{${}^2$ Instituut voor Theoretische Fysica, KU Leuven, Celestijnenlaan 200D B-3001 Leuven, Belgium}

{${}^3$ {Racah Institute of Physics, The Hebrew University of Jerusalem,
Jerusalem 91904, Israel}}

{${}^4$ ORFE Department, Princeton University, Princeton, NJ 08544, USA}

{${}^5$ Institute for Theoretical Physics, University of Amsterdam, Amsterdam, 1098XH, The Netherlands}

 \vspace{0.1in}
 
 {\tt blommaert@ias.edu, jkudlerflam@ias.edu, narovlansky@princeton.edu, urbach@ias.edu}
\end{center}
 
 \vspace{0.4in}
 
\begin{abstract}
\noindent We investigate 3d de Sitter axion wormhole which contribute to the no-boundary density matrix. We identify ``cosmological necklace’’ solutions: an infinite series of Euclidean saddles corresponding with repeated bounces. This results in an unbound gravitational entropy, and a divergent path integral. To remedy this, we study the gravitational path integral using a (mostly) Lorentzian lapse contour. Within a minisuperspace steepest-descent analysis, we find that a single necklace dominates, leading to a finite entropy. A crucial element is to take into account an $(a\to -a)$ redundancy in the FLRW path integral, where $a$ is the scale factor. Surprisingly, the dominant solution is not purely Euclidean. Its entropy turns out to be independent of the axion flux, and equals the empty de Sitter entropy. We also study higher-dimensional necklaces, sourced by either an axion flux or by Yang-Mills instantons, and argue for qualitatively similar results: the single necklace solution dominates along a Lorentzian lapse contour.

 \end{abstract}

 \pagebreak
 \setcounter{page}{1}
 \setcounter{tocdepth}{2}
 \tableofcontents

\section{Introduction and summary}\label{sect:1.intro}

Defining the gravitational path integral via summing over all Euclidean geometries and topologies is notoriously subtle, not least of all because of the unboundedness of the Euclidean action, the so-called conformal factor problem \cite{Gibbons:1978ac}, which is inherently off-shell. Despite these challenges, in 2019, two significant pieces of progress in the context of black hole physics came from summing over large classes of Euclidean spacetimes in Jackiw-Teitelboim gravity, a simple 2d model of AdS quantum gravity \cite{sss,Penington:2019kki}. In particular, the inclusion of Euclidean spacetime wormholes elucidates how black holes can evaporate unitarily (following a Page curve) \cite{Penington:2019kki}, and why late time matter correlators near black holes stop decaying \cite{sss,philsolo,disecting}. This progress emphasized the importance of including non-trivial geometries and topologies in the gravitational path integral. Indeed, in \cite{sss}, ``all'' hyperbolic geometries were included.

In this work, we argue that one should not sum over all geometries and topologies, even when they are real Euclidean solutions to the Einstein field equations (such solutions trivially satisfy the KSW condition \cite{Kontsevich:2021dmb, Witten:2021nzp}). There is earlier evidence that summing over all Euclidean geometries does not make much sense. For instance, in AdS$_3$ the sum over all manifolds ending on the holographic torus boundary seems divergent\footnote{See for instance \cite{Maloney:2007ud} and later work.} (see however \cite{Stanford:2025llj}). This seemingly unreasonable output raises questions about the definition of the path integral itself, as summing over all geometries. Our analysis demonstrates a more severe problem. We identify an infinite class of problematic classical solutions where incorporating any such saddle yields unphysical results: their entropy is greater than the Gibbons-Hawking entropy, thus violating the second law of thermodynamics (as each of these saddles asymptotes to empty dS space at infinite time).

Gibbons and Hawking (GH) \cite{Gibbons:1977mu, gibbons1977action} proposed that an observer in the de Sitter (dS) static patch has an entropy $S=A/4\text{G}$, with $A$ the area of the cosmological horizon. As for the case of black holes, there is both a thermodynamic and a path integral derivation. The path integral derivation follows from evaluating the path integral classical action of the sphere, the Wick rotation of dS,
which equals the exponential of the horizon area:
\begin{equation}
   \begin{tikzpicture}[baseline={([yshift=-.5ex]current bounding box.center)}, scale=0.7]
 \pgftext{\includegraphics[scale=1]{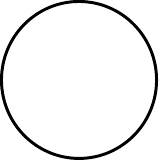}} at (0,0);
  \end{tikzpicture} \quad = e^\frac{A}{4 G}\,.\label{1.1sphere}
\end{equation} 
Rotations analytically continue to static patch time evolution. One is naturally tempted to interpret this sphere as computing some trace of (something like) an identity operator on (something like) the static patch Hilbert space $\mathcal{H}$:\footnote{Subregions in quantum gravity (like the static patch) are difficult to define. Furthermore, even plain QFT on a subregion does not usually have a well-defined trace on some Hilbert space. See for instance \cite{Witten:2018zxz,Witten:2021jzq}. Therefore, it is far from obvious what the relevant Hilbert space and trace would be.}
\begin{equation}
    \begin{tikzpicture}[baseline={([yshift=-.5ex]current bounding box.center)}, scale=0.7]
 \pgftext{\includegraphics[scale=1]{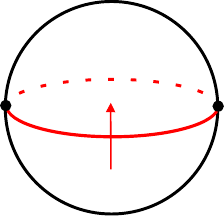}} at (0,0);
    \draw (-0,-1.4) node {\color{red}time};
    \draw (-3,0.1) node {horizon};
  \end{tikzpicture}\quad = e^\frac{A}{4 G}\overset{?}{=}\tr_\mathcal{H}(1)=e^S\,.\label{1.2S}
\end{equation}
Is $A/4\text{G}$ an entropy? Further, largely independent, support for this interpretation came from a Lorentzian calculation that constructed a novel algebra of observables satisfying the Hamiltonian constraint in the dS static patch \cite{Chandrasekaran:2022cip}.\footnote{The other leading order gravitational constraints were imposed in \cite{Klinger:2026tws} leading to further edge mode contributions.} 
Remarkably, they showed that such an algebra has a well-defined notion of von Neumann entropy that equaled the generalized entropy.\footnote{Technically they only showed that generalized entropy differences measure entropy differences between states. A relative state counting interpretation has been discussed in \cite{Akers:2024bel}.} Furthermore, there is suggestive evidence for a state counting interpretation of the sphere path integral at one-loop \cite{Anninos:2020hfj,Anninos:2021ene,Anninos:2026hia,Anninos:2023exn,Maldacena:2024spf,Chen:2025jqm,Law:2026tuk} (see however \cite{Chen:2026boh,Milekhin:2026tbi,Cui:2026bcd,Harlow:2026pwe,Kolchmeyer:2024fly,Narovlansky:2025tpb} for recent tension with this interpretation).

In this work we take the perspective that (much like the Page curve \cite{Page:1993wv} and the late-time correlation function \cite{Maldacena:2001kr,Cotler:2016fpe} in AdS/CFT) $S=A/4\text{G}$ is an external input (for instance, from the Lorentzian CLPW calculation) that helps us better understand the proper definition of the gravitational path integral. In particular, the setting we will be interested in is the no-boundary density matrix \cite{hartle1983wave, Hawking:1983hj, page1986density, hawking1987density, Ivo:2024ill}, and more precisely its trace, which ought to compute the dS entropy according to GH. From this perspective, one views the Gibbons-Hawking proposal as a necessary output of the no-boundary density matrix in a reasonable theory of gravity. We scrutinize this assertion in the discussion \textbf{section \ref{sect5:concl}}.

Based on equation \eqref{1.1sphere}, the Euclidean gravitational path integral may appear to be in good shape at tree level. Unfortunately, as we explain in \textbf{section \ref{sect1.1unentropicsols}}, it is not. Our conclusion will be that the gravitational path integral \emph{cannot} be defined as integrating over all Euclidean spacetimes. We will explore an alternative definition of the gravitational path integral in \textbf{section \ref{sect1.2proposal}}, integrating over (almost) Lorentzian spacetimes. 
Unlike the integral over all Euclidean spacetimes, our definition gives a reasonable prediction for the entropy associated with an observer in more general FLRW cosmologies.

\subsection{The cosmological necklace problem}\label{sect1.1unentropicsols}
In this work, we focus on $d$-dimensional FLRW cosmologies with closed spatial slices, obtained by adding specific matter to Einstein gravity with a positive cosmological constant. Concrete examples include the 3d axion de Sitter wormholes that we will study in \textbf{section \ref{sect2:3daxion}}, higher dimensional axion wormholes of \textbf{section \ref{sect3:5daxion}} and the magnetic wormholes of \textbf{section \ref{sect4:magnetic}}. 

A concrete example of such a spacetime is equation \eqref{2.5Euclideansol}. Adding homogeneous matter to de Sitter space enables a comoving observer to see a greater portion of the spacetime. Depending on the amount of matter stress energy, the observer in this spacetime can have causal access to more than half of the spatial slice, or even a whole spatial slice:
\begin{equation}
    \begin{tikzpicture}[baseline={([yshift=-.5ex]current bounding box.center)}, scale=0.7]
 \pgftext{\includegraphics[scale=1]{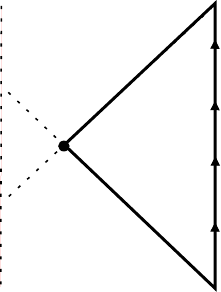}} at (0,0);
    \draw (-0.6,-1.4) node {diamond};
    \draw (2.9,0.1) node {observer};
    \draw (0,-3) node {1. restricted causal access};
  \end{tikzpicture} \qquad \qquad\qquad
    \begin{tikzpicture}[baseline={([yshift=-.5ex]current bounding box.center)}, scale=0.7]
 \pgftext{\includegraphics[scale=1]{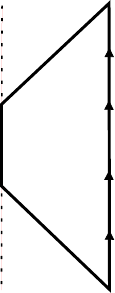}} at (0,0);
    \draw (-2.1,0.1) node {diamond};
    \draw (2,0.1) node {observer};
    \draw (0,-3) node {2. complete causal access};
  \end{tikzpicture} \label{1.4causalacces}
\end{equation}
For the present argument, we consider scenario 1, which limits (as the matter density approaches zero) to dS space. In this case, an observer has a causal horizon. 

For these theories, we study the no-boundary state for an observer existing in the corresponding matter density sector.
The analog of the dS sphere solution, for the axionic and Yang-Mills matter which we consider in this work, would be a Wick rotation of the aforementioned FLRW solutions, of the type 
\begin{equation}
    \d s^2=N^2\d \tau^2+a(\tau)^2\d \Omega_{d-1}^2\,,\quad 0\leq \tau\leq 1,\label{1.3metric}
\end{equation}
with $N$ the Euclidean lapse. To investigate the entropy, as in \eqref{1.2S}, one takes $a(0)=a(1)$, which results in a compact geometry (the unusual conditions for matter configurations to result in such Euclidean wormhole solutions are discussed in section \ref{sect4:magnetic}). We can draw these wormhole solutions as follows
\begin{equation}
    \begin{tikzpicture}[baseline={([yshift=-.5ex]current bounding box.center)}, scale=0.7]
 \pgftext{\includegraphics[scale=1]{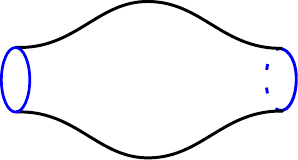}} at (0,0);
    \draw (-3.2,0.1) node {\color{blue}glue};
  \end{tikzpicture} \quad = e^{-I_0} \neq e^\frac{A}{4 G},\label{1.5worm}
\end{equation}
where the two blue ends are glued together to produce a closed Euclidean spacetime and $I_0<0$ is the solution's on-shell action. In the empty dS limit, the blue region pinches off, reproducing the sphere. Unlike the sphere, the wormhole action has no (obvious) interpretation as an extremal area, or the area of the boundary of a stationary observer's causal diamond. Nevertheless, with motivations from \cite{witten2024background} it was suggested in \cite{Blommaert:2025bgd} that $-I_0$ could be interpreted as the gravitational entropy associated with the observer's causal diamond (or, a Cauchy slice of the diamond):\footnote{The analogy with the sphere is not completely obvious. Indeed, it is not obvious if an analytic continuation of a Lorentzian time in the causal patch of an observer in this spacetime maps nicely to a Euclidean time coordinate on this wormhole. However, independent of this confusion, arguments were presented in \cite{Blommaert:2025bgd} that \eqref{1.5worm} does actually compute the observer's entropy. A perturbative calculation supporting this proposal, along the lines of CLPW \cite{Chandrasekaran:2022cip}, would be extremely valuable.}
\begin{equation}
     \begin{tikzpicture}[baseline={([yshift=-.5ex]current bounding box.center)}, scale=0.7]
 \pgftext{\includegraphics[scale=1]{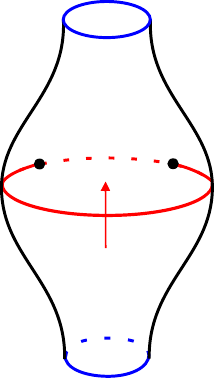}} at (0,0);
    \draw (0,-1.4) node {\color{red}time};
  \end{tikzpicture} \quad = \,e^{-I_0}\overset{?}{=}\tr_\mathcal{H}(1)=e^S\,.
\end{equation}

So far, the picture seems reasonable. The issue, however, is that we can easily construct more dominant saddles by gluing $k\in \mathbb{Z}$ copies of the wormhole solutions,
\begin{equation}
    \begin{tikzpicture}[baseline={([yshift=-.5ex]current bounding box.center)}, scale=0.7]
 \pgftext{\includegraphics[scale=1]{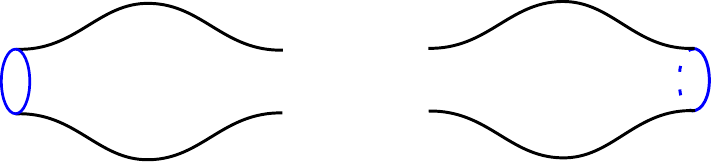}} at (0,0);
    \draw (-6.7,0.1) node {\color{blue}glue};
    \draw (0,0.1) node {\dots};
    \draw (0,-1.5) node {$k$ copies};
  \end{tikzpicture}\quad = e^{-k I_0}\,.\label{1.7oscillate}
\end{equation}
These solutions are also closed Euclidean saddles with the same boundary conditions, only with a different lapse $N=k N_0$, with $N_0$ the lapse of the original wormhole solution.
Being simple covers of the original solution, no local criterion can allow the original solution but exclude its covers. Considering all these solutions as contributions to the no-boundary trace, the entropy of each is $k \cdot |I_0|$. Thus, any $k>1$ cover is more dominant than the wormhole. Worse, there is no dominant saddle and the entropy diverges. We refer to this as the necklace problem. What makes this setup stand out is that the divergence comes from a sum over real classical \textbf{solutions}, rather than field configurations.
We therefore conclude that summing over all Euclidean classical solutions, and excitations around them, is not a reasonable definition of a gravitational path integral. 

In the remainder of this work we advance
an alternative definition of the gravitational path integral, one that leads to a convergent answer for the sum over geometries and that agrees with GH in the empty dS limit. We note that the \textbf{necklace problem} persists in the empty dS case if one allows for disconnected manifolds: a Euclidean path integral sums over multiple copies of the Euclidean sphere:\footnote{See also for instance \cite{Klebanov:1988eh,halliwell1989multiple}. }
\begin{equation}
    \begin{tikzpicture}[baseline={([yshift=-.5ex]current bounding box.center)}, scale=0.7]
 \pgftext{\includegraphics[scale=1]{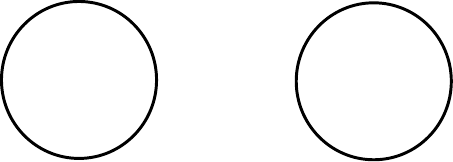}} at (0,0);
    \draw (0,0.1) node {\dots};
    \draw (0,-1.5) node {$k$ copies};
  \end{tikzpicture}\quad = e^{-k I}
\end{equation}
which can be thought of as a limit where the wormhole throat becomes small.
This is inconsistent with GH. See \textbf{section \ref{sec:pure_grav_3d}}. Our \textbf{proposal rules out necklaces} with $k>1$.

\subsection{Resolution: Lorentzian lapses}\label{sect1.2proposal}
Instead of a local criterion for on-shell solutions, we will consider an off-shell prescription for the no-boundary path integral that will yield global constraints on possible saddles. To go off-shell, we will consider the gravitational path integral in the ADM formalism in minisuperspace variables; see \textbf{section \ref{sect2:3daxion}}. The dynamical fields are $N$ and $a$ (and the conjugate momentum $p$ of $a$) from \eqref{1.3metric}. 
We consider quantum gravity defined by the following \textbf{Lorentzian lapse} contour:\footnote{We impose initial and final boundary conditions $a=a_\text{max}$ at $\tau=0$ and $\tau=1$. For real $N$ the metric \eqref{1.3metric} is Euclidean between the initial and final slice $0\leq \tau\leq 1$. If $N=\i \mathbb{R}$ the metric preparing the state is Lorentzian between $0<\tau<1$. Lorentzian lapses means that we integrate over Lorentzian path integral preparations. The dS sphere has a Euclidean lapse (real $N$). The invariant statement is whether the integral of $\d s$ between initial and final Cauchy slices is real (Euclidean lapse), or purely imaginary (Lorentzian lapse).}
\begin{equation}
    \boxed{N=\varepsilon+\i \mathbb{R}\,}
\end{equation}
Lorentzian lapse contours have a long history in quantum cosmology \cite{Teitelboim:1981ua,Teitelboim:1983fh,Teitelboim:1983fk,Vilenkin:1986cy,Linde:1983mx,Halliwell:1988ik,Halliwell:1989vu,Halliwell:1990tu,Halliwell:1989dy,Feldbrugge:2017kzv,Feldbrugge:2017fcc,Feldbrugge:2017mbc,Marolf:1996gb,Marolf:2022ybi,Loges:2022nuw}; see \cite{Lehners:2023yrj} for a review.\footnote{In general, the lapse contour is a question of active investigation, see for instance \cite{DiazDorronsoro:2017hti,Feldbrugge:2017kzv,DiTucci:2019bui,Banihashemi:2024aal,Dittrich:2024awu,Blommaert:2025bgd}. The real axis is natural in quantum mechanics and gives reasonable answers, however it is not obviously equivalent to calculations for instance in 2d quantum gravity, and in matrix models \cite{Anninos:2026eqv,Moore:1991ir}. We thank Dio Anninos and Nathan Seiberg for discussions. Timelike Liouville might provide a rigid framework to make progress on this \cite{Anninos:2024iwf,Anninos:2025fer}.}

In order to study the effects of this contour, it will prove useful to first integrate out $a(\tau)$ and obtain an effective action $I(N)$ for the no-boundary trace. We will either do it exactly when possible, as in \textbf{section \ref{sect2.2exact}}, or by saddle point when not possible. This action will have saddles at $N=k\cdot N_0$, representing necklaces.
We will see in \textbf{section \ref{sect2:3daxion}} that there are two important and topologically distinct classes of solutions for $a(\tau)$. They arise because the metric depends only on the combination $a^2$. This means that if we compute the density matrix of the universe to propagate from a slice with metric $a^2\d\Omega_d^2$ to itself, we have to sum two propagators in the minisuperspace quantum mechanics (where $a$ is a particle with potential $V(a)$, see equation \eqref{potential}):
\begin{equation}
    \mathcal{G}_N(a\rvert a)+\mathcal{G}_N(a\rvert -a)\,.\label{1.10G+G}
\end{equation}
The other combinations $\mathcal{G}_N(-a\rvert -a)$ and $\mathcal{G}_N(-a\rvert a)$ produce a factor of two. In other words, we are treating $\bf{a\leftrightarrow-a}$ as a \textbf{gauge redundancy}. We now describe the configurations $a(\tau)$ that contribute to each propagator.

\subsubsection*{Configuration 1: Tunneling solutions}

Consider first the configurations that contribute to $\mathcal{G}_N(a\rvert -a)$. We will conclude that these configurations capture saddles with odd values of $k=2n+1$. In terms of a (typical) FLRW potential $V(a)$ (see \textbf{section \ref{sect2.1setup}} for an explicit example) the simplest solution (with $k=1$ or $N=N_0$) looks as follows:\footnote{This picture for $V(a)$ is most clear in 3d (see section \ref{sect2:3daxion}). It extends to odd spacetime dimensions in a rather straightforward manner (see section \ref{sect3:5daxion}). For even spacetime dimensions, this symmetry between $a$ and $-a$ is less obvious (see section \ref{sect3:5daxion}), but persists in (what we will call) kinetic gauge.}
\begin{equation}
    \begin{tikzpicture}[baseline={([yshift=-.5ex]current bounding box.center)}, scale=0.7]
    \pgftext{\includegraphics[scale=1]{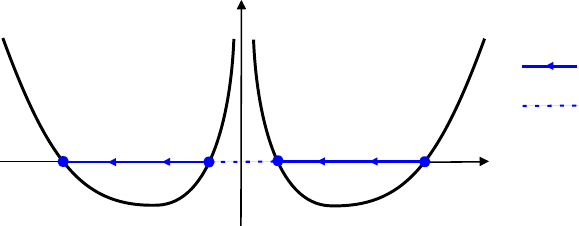}} at (0,0);
    \draw (6.3,0.9) node {\color{blue}Euclidean};
    \draw (6.4,0.2) node {\color{blue}Lorentzian};
    \draw (2.6,-1.2) node {\color{blue}$a$};
    \draw (-4.3,-1.2) node {\color{blue}$-a$};
    \draw (0.5,-0.35) node {\color{blue}$a_\text{min}$};
    \draw (0,1.65) node {$V(a)$};
    \end{tikzpicture} \label{1.11potential}
\end{equation}
The universe starts at spatial size $a$ (we consider $a=a_\text{max}$, the thickest part of the necklace in equation \eqref{1.5worm}) and decreases in size, until it reaches $a_\text{min}$. At this point, the Euclidean solution tunnels through a Lorentzian region and goes through $a=0$. Then $a$ becomes negative and grows in absolute value, first in Lorentzian space, and then in Euclidean space until the size of the universe again reaches $\abs{a}=a_\text{max}$. 

Summarizing: the geometry $a(\tau)$ for $k=1$ ($N=N_0$) looks like:\footnote{The two slices where $\abs{a}=a_\text{max}$ are identified such as to produce a closed topology, in the spirit of equation \eqref{1.5worm}. $a_\text{max}$ and $a_\text{min}$ are surfaces of time-reflection symmetry (and zero extrinsic curvature) where one could glue geometries together, or continue smoothly to Lorentzian signature spacetimes.}
\begin{equation}
    \begin{tikzpicture}[baseline={([yshift=-.5ex]current bounding box.center)}, scale=0.7]
    \pgftext{\includegraphics[scale=1]{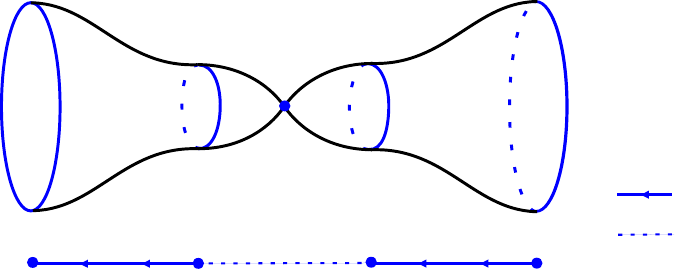}} at (0,0);
    \draw (7.1,0.-0.9) node {\color{blue}Euclidean};
    \draw (7.2,-1.6) node {\color{blue}Lorentzian};
    \draw (3.4,-2.6) node {\color{blue}$a$};
    \draw (-5.2,-2.6) node {\color{blue}$-a$};
    \draw (0.6,-2.6) node {\color{blue}$a_\text{min}$};
    \draw (-2.4,-2.6) node {\color{blue}$-a_\text{min}$};
    \draw (-6.5,0.4) node {\color{blue}glue};
    \draw (-3.15,1.95) node {$a(\tau)$};
    \end{tikzpicture} \label{1.12dominant}
\end{equation}
For 3d axion wormholes, one can show that this solution contributes to $\mathcal{G}(a\rvert -a)$ by analyzing the exact quantum propagator (see \cite{halliwell1989multiple} for a related calculations). We avoid the big-bang singularity at $a=0$ by going in the complex $a$ plane, resulting in a bra-ket wormhole \cite{Chen:2020tes,Fumagalli:2024msi}. For higher necklaces $k>1$, we paste $2n+1$ copies of this spacetime together. 

The complex action $I_{a\to-a}(N)=-\log \mathcal{G}(a\rvert-a)$ schematically has the following structure:
\begin{equation}
    \begin{tikzpicture}[baseline={([yshift=-.5ex]current bounding box.center)}, scale=0.7]
    \pgftext{\includegraphics[scale=1]{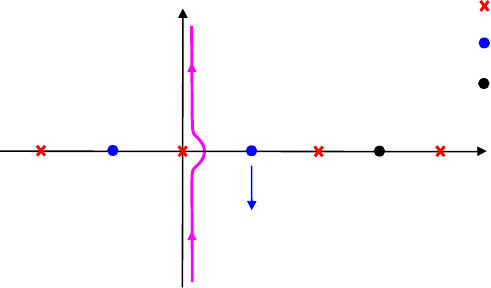}} at (0,0);
    \draw (4.9,2.4) node {\color{red}pole};
    \draw (6.75,1.7) node {\color{blue}contributing saddle};
    \draw (7.25,1) node {non-contributing saddle};
    \draw (3.9,-0.8) node {$N$};
    \draw (1.6,-1.5) node {\color{blue}dominant saddle};
    \draw (0.7,1.75) node {\color{mygenta}$N=\varepsilon+\i\mathbb{R}$};
    \draw (-2.2,0.4) node {\color{blue}$-N_0$};
    \draw (0.1,0.4) node {\color{blue}$N_0$};
    \draw (2.3,0.4) node {$3N_0$};
    \draw (-2.9,2) node {$I_{a\to-a}(N)$};
    \end{tikzpicture}\label{1.13contours1}
\end{equation}
A \textbf{Picard-Lefschetz} analysis crucially shows that only the saddles $k\leq 1$ contribute. This avoids the problematic saddles. The geometry pictured in \eqref{1.12dominant} is the leading contribution to the gravitational path integral, and predicts an entropy $S=-I$ that agrees with GH in the empty dS limit. The key feature that makes us pick up the usual GH saddle ($k=1$) is the presence of a \textbf{pole for} $\bf{N\to0}$ in the propagator $\mathcal{G}_N(a\rvert -a)$. This arises in the quantum mechanics because it takes infinite energy to go from $a\to -a$ in no time (see \textbf{section \ref{sect2.2exact}}). This is the reason why the Lorentzian contour $N=\varepsilon+\i \mathbb{R}$ can pick up a positive entropy saddle, even though Lorentzian metrics have entirely imaginary actions. Positive entropy requires this pole. Fortunately, this pole is universal. See \textbf{section \ref{subsect2.5.4gen}}.

Note that solutions with negative $e^{\varphi} \equiv a<0$ also appear to contribute in the timelike Liouville path integral \cite{Harlow:2011ny}, along the Hankel contour \cite{Usciati:2025cdn}. In timelike Liouville, the rigid framework of 2d CFT can be used to better understand the rules of the gravitational path integral. In this setup, 2d CFT naively therefore seems to agree that $a<0$ should be taken seriously.

\subsubsection*{Configuration 2: Standard solutions}

Next, we consider the configurations which contribute to $\mathcal{G}_N(a\rvert a)$. We will find that these configurations capture saddles with even $k=2n$. These consist of $2n$ copies of the spacetime in equation \eqref{1.12dominant}, glued together. In terms of the potential $V(a)$ \eqref{1.11potential}, the universe bounces back and forth $n$ times, ending and starting at an identical value of $a$. The case $k=0$ describes a universe with total elapsed time $N=0$ (static) with size $a=a_\text{max}$. The complex structure of $I_{a\to a}(N)=-\log \mathcal{G}(a\rvert a)$ with a contour $N=\varepsilon+\i \mathbb{R}$ looks as follows:
\begin{equation}
    \begin{tikzpicture}[baseline={([yshift=-.5ex]current bounding box.center)}, scale=0.7]
    \pgftext{\includegraphics[scale=1]{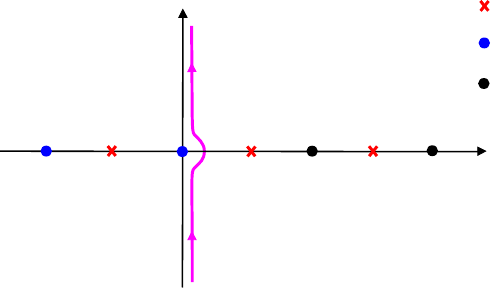}} at (0,0);
    \draw (4.9,2.4) node {\color{red}pole};
    \draw (6.75,1.7) node {\color{blue}contributing saddle};
    \draw (7.25,1) node {non-contributing saddle};
    \draw (3.9,-0.8) node {$N$};
    \draw (0.7,1.75) node {\color{mygenta}$N=\varepsilon+\i\mathbb{R}$};
    \draw (-3.5,0.4) node {\color{blue}$-2N_0$};
    \draw (1.15,0.4) node {$2N_0$};
    \draw (3.2,0.4) node {$4N_0$};
    \draw (-3.1,2) node {$I_{a\to a}(N)$};
    \end{tikzpicture}\label{1.14contours2}
\end{equation}
The steepest descent contours show that only those saddles with $n<1$ contribute. We stress that the problem of unphysical saddles with a Euclidean action that is unbounded from below is already there within this more standard $\mathcal{G}(a\rvert a)$ calculation. A Lorentzian $N=\varepsilon+\i \mathbb{R}$ contour avoids this divergence.

\subsubsection*{Conclusion}

Thus, defining the gravitational path integral through the contour $N=\varepsilon+\i \mathbb{R}$ rules out the problematic saddles ($k>1$) with unbound actions. The \textbf{Gibbons-Hawking saddle dominates}. This is consistent with the prediction $S=-I$ for the entropy associated with an FLRW observer, and agrees with GH in the empty dS limit. This suggests that $N=\varepsilon+\i \mathbb{R}$ is the appropriate contour to define the gravitational path integral, from an observer's perspective.

As we highlight more in the discussion \textbf{section \ref{sect5:concl}}, another possible Lorentzian contour would be $N=-\varepsilon+\i \mathbb{R}$ or $N = \i \mathbb{R}^+$. These contours only pick up the saddles with $k\leq 0$ and crucially not $k=1$.
This suggests $k=0$ as the dominating gravitational ``saddle'' as predicted by the contour $N=-\varepsilon+\i \mathbb{R}$.
Since the only non-trivial saddles contribute non-perturbatively suppressed terms in $G$, one might call it a ``Vilenkin-type contour'' \cite{vilenkin1982creation,Halliwell:1988ik,Vilenkin:1984wp, Vilenkin:1986cy, Vilenkin:1987kf, Linde:1983mx}.

A comment on fluctuations is in order. The conformal mode has wrong-sign fluctuations for $\text{Re}\,N>0$. A purely Lorentzian lapse contour avoids this problem. Because of the positive regulator, the contour $N=\varepsilon+\i\mathbb{R}$, unfortunately, does not. Equivalently: the GH saddle, which dominates the path integral, still suffers from a conformal mode problem. On the bright side, fluctuations of matter fields \emph{are} stable for $\text{Re}\,N>0$. On the Vilenkin contour, the conformal mode problem is resolved but all matter fluctuations are unstable. We comment more on ``allowability'' of these saddles in the discussion \textbf{section \ref{sect5:concl}}.

Throughout, we will rely on the following standard statement of Picard-Lefschetz theory \cite{Witten:2010cx,Feldbrugge:2017kzv}. For an integral of $e^{-I(N)}$ with $I(N)$ holomorphic away from isolated singularities, the integration contour $\mathcal{C}$ is deformed to a sum of steepest descent contours $\mathcal{D}_\sigma$ (Lefschetz thimbles), attached to saddle points $\sigma$:
\begin{equation}
    \int_\mathcal{C}\d N\, e^{-I(N)}=\sum_\sigma n_\sigma \int_{\mathcal{D}_\sigma}\d N\, e^{-I(N)}\,.
\end{equation}
Here, the integer $n_\sigma$ is the intersection number of $\mathcal{C}$ with the steepest ascent contour $\mathcal{A}_\sigma$ of the saddle $\sigma$. A saddle contributes when $n_\sigma \neq 0$. Both $\mathcal{D}_\sigma,\mathcal{A}_\sigma$ have constant $\text{Im}\,I(N)$ (constant phase). 

\subsection{Structure}
In the remainder of this article, we will work out detailed examples. This work is organized as follows.

In \textbf{section \ref{sect2:3daxion}} we investigate 3d axion-dS wormholes \cite{Myers:1988sp, Aguilar-Gutierrez:2023ril}. Following \cite{halliwell1989multiple} we compute the propagators $\mathcal{G}_N(a\rvert \pm a)$ via the minisuperspace quantum mechanics, and compare with classical solutions. We then perform a steepest descent analysis. We find that the $k=1$ classical solution dominates for the contour $N=\varepsilon+\i \mathbb{R}$. The $k=0$ saddle dominates for the contour $N=-\varepsilon+\i \mathbb{R}$. One very interesting but puzzling fact is that the resulting $k=1$ action is independent of the amount of axion flux $q$:
\begin{equation}
    \boxed{-I=\frac{\pi}{2G}\overset{?}{=}S\quad q\text{-independent}\,}\label{univ}
\end{equation}
This equals $A/4G$ for the empty dS space. This gives a clear and sufficiently simple target for Lorentzian algebraic calculations to reproduce (or disprove). We leave a further investigation of this fact, including whether this generalizes to higher dimensions, to future work.

In \textbf{section \ref{sect3:5daxion}} we study higher dimensional axion wormholes numerically. These numerics take place in what we call kinetic gauge. In \textbf{appendix \ref{app:5dproper}} we study 5d axion wormholes in what we call proper-time gauge, analytically and numerically.

In \textbf{section \ref{sect4:magnetic}} we construct Euclidean wormhole solutions for Einstein gravity coupled to SO$(d-1)$ Yang-Mills theory. These SO$(d-1)$ \textbf{magnetic wormholes} have a cosmological necklace problem with identical resolution as the axion case.

Finally, in the discussion \textbf{section \ref{sect5:concl}} we take stock of the two Lorentzian lapse contours $N=\pm \varepsilon+\i \mathbb{R}$.

\section{Three dimensional axion necklaces}\label{sect2:3daxion}
This section is the core of this paper. We work out in detail the propagators $\mathcal{G}_N(a\rvert \pm a)$ for 3d axion-dS wormhole solutions \cite{Aguilar-Gutierrez:2023ril,halliwell1989multiple} and find the general structure claimed in section \ref{sect1.2proposal}. In \textbf{section \ref{sect2.1setup}}, we explain the details of our setup (namely, the definition of the propagators $\mathcal{G}_N(a\rvert \pm a)$ and which of their properties we will investigate). In \textbf{section \ref{sect2.2exact}} (within the minisuperspace approximation), we exactly evaluate the gravitational propagators $\mathcal{G}_N(a\rvert \pm a)$. We study the classical limit of the exact answer and find the structure of saddles and steepest descent contours as claimed in equation \eqref{1.13contours1} and \eqref{1.14contours2}.

The exact quantum propagator is powerful, because it implicitly already makes a selection in which configurations contribute to the gravitational path integral. We clarify this point in \textbf{section \ref{sect2.4necklasesemi}}, where we study the classical solutions of the equations of motion, and identify which of those solutions actually do contribute in the quantum propagators $\mathcal{G}_N(a\rvert \pm a)$. Two features which we will confirm are that the odd saddles $k=2n+1$ appear as saddles of (the effective action for) $\mathcal{G}_N(a\rvert -a)$, as discussed around \eqref{1.11potential}, and that the even saddles $k=2n$ appear only in $\mathcal{G}_N(a\rvert a)$. Which of these saddles actually contributes is then determined by the steepest descent analysis. On the contour $N=\varepsilon+\i\mathbb{R}$ only $k=1$ and the (suppressed) $k<0$ saddles contribute in the odd sector, and only $k\leq 0$ in the even sector. We furthermore show that the Lorentzian tunneling in the odd sectors as depicted in equation \eqref{1.12dominant} can be deduced from the exact propagator $\mathcal{G}_N(a\rvert \pm a)$. 

As this analysis contains several elements, we first scrutinize the classical solutions for empty dS in \textbf{section \ref{sec:pure_grav_3d}}, where a contour $N=\varepsilon+\i \mathbb{R}$ results in a single sphere as the dominant saddle. In \textbf{section \ref{subsect2.5.4gen}} we briefly discuss a generalization of part of our argument to more general FLRW potentials $V(a)$. In \textbf{section \ref{sect:3.3}} we provide an intuitive argument for why the on-shell action of our 3d axion wormholes is $q$-independent, matching the empty dS cosmological horizon area \eqref{univ}: the contour which computes the on-shell action may be deformed to late times - where any FLRW solution will approach empty dS.

\subsection{Setup}\label{sect2.1setup}
Following \cite{Myers:1988sp,halliwell1989multiple} we consider 3d axion-dS wormholes. See \cite{Aguilar-Gutierrez:2023ril} for background. Consider the Euclidean action (up to boundary terms) with Hubble constant set to $H=1$: 
\begin{equation}
    I= -\frac{1}{16\pi G}\int \d^3 x\sqrt{g}(R-2)+\frac{1}{4}\int\d^3 x\sqrt{g} H^{\mu \nu}H_{\mu\nu}\,,\quad H_{\mu\nu}=Q\, \eps_{\mu\nu}(S^2)\,,\quad Q^2 =  \frac{q^2}{16\pi G}\, \label{2.1ac}
\end{equation}
where $\eps_{\mu\nu}(S^2)$ is the volume form of the spatial sphere.
Consider furthermore the FLRW ansatz \eqref{1.3metric}:
\begin{equation}
\label{2.1FLRW_coord}
 \d s^2=N^2\d \tau^2+ a(\tau)^2\d \Omega_2\,,\quad \d \Omega_2=\d\theta^2+\sin^2(\theta)\d \phi^2\,,\quad 0\leq \tau\leq 1\,.
\end{equation}
Throughout this section, we will study the gravitational path integral with boundary conditions where one fixes the induced boundary geometry on the initial and final Cauchy slices:
\begin{equation}
    \d s^2_\text{boundary}\rvert_{\tau=0}=a_0^2 \d \Omega_2^2\,,\quad \d s^2_\text{boundary}\rvert_{\tau=1}=a_1^2 \d \Omega_2^2\,.\label{2.10boudnarycond}
\end{equation}
The range of the time variable $0\leq \tau\leq 1$ is convention, placing all the dynamical data about the spacetime in the fields $N$ and $a(\tau)$. The action \eqref{2.1ac} evaluated on FLRW spacetimes \eqref{2.1FLRW_coord} with $0\leq \tau\leq 1$ becomes:
\begin{equation}
    \boxed{I=\frac{1}{\text{G}}\int_0^1\d \tau \bigg\{ -\frac{1}{2N}\dot{a}^2+\frac{N}{2}\bigg(-1+a^2+\frac{q^2}{4}\frac{1}{a^2}  \bigg) \bigg\}\,}\label{2.8action}
\end{equation}
The Friedmann equation is obtained by variation with respect to the lapse function $N$:
\begin{align}
\frac{1}{N^2}\frac{\dot{a}^2}{a^2} -\frac{1}{a^2} +1 + \frac{q^2}{4 }\frac{1}{a^4}=0\,.\label{2.4friedmann}
\end{align}
The solutions are:
\begin{equation}
    a^2=\frac{1-\sqrt{1-q^2}}{2}+\sqrt{1-q^2}\cos(N \tau)^2\,.\label{2.5Euclideansol}
\end{equation}
Notice that the total Euclidean proper time between the two ends of the spacetime equals $N$. Rotating $\tau=\i T$ we obtain Lorentzian expanding dS space:
\begin{equation}
 a^2=\frac{1-\sqrt{1-q^2}}{2}+\sqrt{1-q^2}\cosh( NT)^2\,.\label{2.3}
\end{equation}
These asymptote to exponentially expanding 3d dS space, as expected. Indeed, the axion term in \eqref{2.4friedmann} becomes subdominant for $a\to\infty$. The maximal size of the Euclidean solution and the minimal size of the Lorentzian solution occur at the time-reflection symmetric point from which Lorentzian expansion starts. This maximal size equals:
\begin{equation}
    a_\text{max}^2=\frac{1+\sqrt{1-q^2}}{2}\,.\label{2.6max}
\end{equation}
The solutions oscillate in Euclidean signature. Variation with respect to $a$ of \eqref{2.8action} produces the second Friedmann equation:
\begin{equation}
    \frac{1}{N^2}\ddot{a}+a-\frac{q^2}{4}\frac{1}{a^3}=0\,.\label{2.9aeq}
\end{equation}
Observe that solving the $N$ EOM \eqref{2.4friedmann} guarantees a solution for the $a$ EOM (acting with $\dot{\cdots}$ on the $N$ EOM produces the $a$ EOM). But solving the $a$ EOM does not guarantee a solution for the $N$ EOM. 

A path integral with action \eqref{2.8action}, initial value $a_0$ and final value $a_1$ decomposes into minisuperspace quantum mechanics propagators as follows:
\begin{equation}
   \int_\mathcal{C}\d N \int_{a_0}^{a_1} \mathcal{D}a\,e^{-I}=\int_\mathcal{C}\d N\,\mathcal{G}_N(a_0\rvert a_1)\,.\label{2.11GI}
\end{equation}
Here $\mathcal{G}_N(a_0\rvert a_1)$ is a propagator of the $a$ quantum mechanics describing propagation from $a_0$ to $a_1$ with Euclidean time $N$:
\begin{equation}
    \mathcal{G}_N(a_0\rvert a_1)=\bra{a_1}e^{-\frac{N}{G} H_\text{WDW}}\ket{a_0}\,,\quad H_\text{WDW}=E=\frac{p^2}{2}+V(a)\,,\quad \boxed{2V(a)=-1+a^2+\frac{q^2}{4}\frac{1}{a^2}\,}\label{potential}
\end{equation}
The Hamiltonian for this minisuperspace quantum mechanics is the gravitational WDW Hamiltonian \cite{DeWitt:1967yk}: $H_\text{WDW}$. This Hamiltonian describes the propagation of a particle (the universe) in the potential $V(a)$:
\begin{equation}
    \begin{tikzpicture}[baseline={([yshift=-.5ex]current bounding box.center)}, scale=0.7]
    \pgftext{\includegraphics[scale=1]{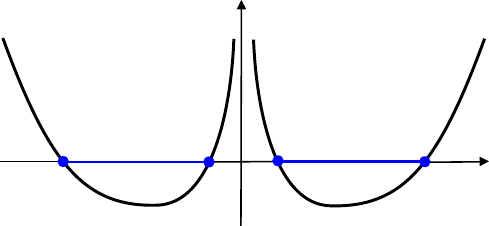}} at (0,0);
    \draw (-5.1,-0.8) node {$E=0$};
    \draw (3.7,-1.2) node {\color{blue}$a_\text{max}$};
    \draw (-3.9,-1.2) node {\color{blue}$-a_\text{max}$};
    \draw (1.2,-0.35) node {\color{blue}$a_\text{min}$};
    \draw (0.7,1.65) node {$V(a)$};
    \end{tikzpicture} 
\end{equation}
Note that $a$ can take values on all of $\mathbb{R}$. Here, $p$ is the Euclidean momentum conjugate to $a$: $p=-\dot{a}/N$. The $N$ equation of motion (the Hamiltonian constraint) is saying that the universe has vanishing energy:
\begin{equation}
    H_\text{WDW}=\frac{p^2}{2}+V(a)=0\,.
\end{equation}
This is the usual statement that ``the gravitational Hamiltonian vanishes''. This is a consequence of the fact that time reparameterizations are redundancies in gravity. One such reparameterization changes the value of $N$. This translates in minisuperspace into the fact that, in \eqref{2.8action}, $N$ shows up as a Lagrange multiplier (after going to first order variables, see \eqref{2.23actionlor}), which indeed enforces that $H_\text{WDW}=0$.

\subsubsection*{Gauged quantum mechanics}

To investigate a gravitational path integral subject to boundary conditions \eqref{2.10boudnarycond}, we should take into account the important fact that $a$ and $-a$ in the quantum mechanics describe physically equivalent spatial metrics $a^2\d \Omega_2^2$. As such, the gravitational path integral decomposes into a sum of \emph{four} propagators, integrated over elapsed time:
\begin{align}
   &\rho_\text{no-boundary}(a_0^2\rvert a_1^2)= \quad \begin{tikzpicture}[baseline={([yshift=-.5ex]current bounding box.center)}, scale=0.7]
 \pgftext{\includegraphics[scale=1]{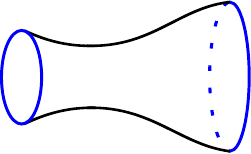}} at (0,0);
    \draw (2.6,0) node {\color{blue}$a_1^2$};
    \draw (-2.65,0) node {\color{blue}$a_0^2$};
    \draw (0,0) node {$\cdots$};
  \end{tikzpicture} \nonumber
   \\&\qquad = \int_\mathcal{C}\d N\,\mathcal{G}_N(a_0\rvert a_1)+\int_\mathcal{C}\d N\,\mathcal{G}_N(a_0\rvert -a_1)+\int_\mathcal{C}\d N\,\mathcal{G}_N(-a_0\rvert a_1)+\int_\mathcal{C}\d N\,\mathcal{G}_N(-a_0\rvert -a_1)\label{2.15rho}
\end{align}
The question which we are investigating in this work is what the correct gravitational contour $\mathcal{C}$ for the lapse integral is. As explained in the introduction section \ref{sect1.1unentropicsols}, the naive Euclidean gravitational path integral would amount to integrating over positive real lapses $0\leq N<\infty$. We now detail the statement made around equation \eqref{1.7oscillate}: this Euclidean contour leads to a divergent gravitational amplitude.

For the gravitational path integral to produce a prediction for the entropy associated with the causal patch of an observer in the spacetime \eqref{2.3},\footnote{For some observer at $\theta=0$ the causal horizon at the time reflection symmetric slice is located at \cite{Blommaert:2025bgd}
\begin{equation}
 \theta_h=\sqrt{\frac{2}{1+\sqrt{1-q^2}}} \ 
 K\bigg(\frac{1-\sqrt{1-q^2}}{1+\sqrt{1-q^2}}\bigg)>\frac{\pi}{2}\,.
\end{equation}
As indicated already in the Penrose diagram equation \eqref{1.4causalacces}, this means this observer has causal access to more than half of the time reflection symmetric global slice.} as explained in section \ref{sect1.1unentropicsols}, we are interested in computing a closed gravitational path integral. We will do this semiclassically. The classical closed solutions are of the type shown in equation \eqref{1.7oscillate}. They are obtained by gluing the Euclidean solution \eqref{2.5Euclideansol} at matching zero extrinsic curvature surfaces, where $a^2=a_\text{max}^2$.\footnote{Alternatively, we could glue at $a^2=a_\text{min}^2$, as shown in equation \eqref{1.7oscillate}. All calculations in this section could be repeated for $\rho_\text{no-boundary}(a_\text{min}^2\rvert a_\text{min}^2)$ with identical results.} Imposing the boundary conditions:
\begin{equation}
    a^2\rvert_{\tau=0}=a^2\rvert_{\tau=1}=a^2_\text{max}\,,\label{2.16bc}
\end{equation}
on the classical solutions \eqref{2.5Euclideansol}, determines the set of classical solutions for $N$ and $a$ completely:
\begin{equation}
    N=k\pi\,,\quad  a^2=\frac{1-\sqrt{1-q^2}}{2}+\sqrt{1-q^2}\cos(k \pi \tau)^2\,,\quad 0<\tau<1\,,\quad k\in\mathbb{Z}\,.\label{2.17solfinal}
\end{equation}
These solutions look like necklaces, once we do the gluing of the $\tau=0$ and $\tau=1$ zero extrinsic curvature slices (as was required to obtain a solution which contributes to the closed gravitational path integral). The Euclidean lapse contour $0\leq N<\infty$ would thus semiclassically correspond to
\begin{equation}
    \mathcal{Z}_\text{no-boundary}= \sum_{k=0}^\infty\qquad 
    \begin{tikzpicture}[baseline={([yshift=-.5ex]current bounding box.center)}, scale=0.7]
 \pgftext{\includegraphics[scale=1]{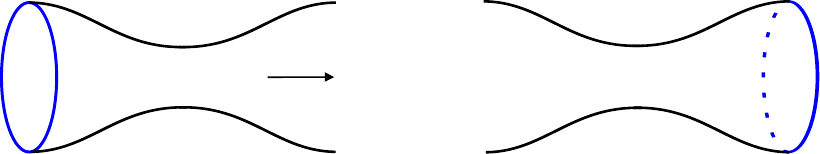}} at (0,0);
    \draw (-7.7,0.1) node {\color{blue}glue};
    \draw (0,0.1) node {\dots};
    \draw (0,-1.8) node {$k$ copies};
    \draw (-6.5,-1.7) node {\color{blue}$\tau=0$};
    \draw (6.5,-1.7) node {\color{blue}$\tau=1$};
    \draw (6.5,1.8) node {\color{blue}$\abs{a_\text{max}}$};
    \draw (-6.5,1.8) node {\color{blue}$\abs{a_\text{max}}$};
    \draw (-1.85,-0.4) node {$\tau$};
    \draw (-2.6,1.5) node {$a(\tau)$};
  \end{tikzpicture}
\end{equation}
The Euclidean on-shell action for each necklace \eqref{2.17solfinal} is negative and equals: 
\begin{equation}
    I_\text{Euclidean}=-\frac{k\pi (1-q)}{2G}<0\,.
\end{equation}
A Euclidean lapse contour $0\leq N<\infty$ sums over the $k\geq 0$ saddles and leads to a divergent Euclidean path integral. This is a concrete example of the necklace problem as explained in section \ref{sect1.1unentropicsols}. The sum over all Euclidean saddles makes no sense \cite{Fischler:1988ia,Fischler:1989ka, Blommaert:2025bgd}, and hence neither does the Euclidean gravitational path integral (given that physically interesting quantities had better be finite).

\subsubsection*{Plan for the remainder of this section}

We will investigate a definition of the lapse contour $\mathcal{C}$ in equation \eqref{2.11GI} which \emph{does} lead to a finite and reasonable answer for the gravitational path integral; and agrees with the Lorentzian semiclassical prediction \eqref{1.2S} by CLPW for empty dS. The gravitational path integral for closed amplitudes takes semiclassically the following form:\footnote{When treating $a\mapsto-a$ as a $\mathbb{Z}_2$ gauge symmetry, the overall factor $2$ in \eqref{2.20Z} should be replaced by $1/2$. However, such $N$-independent prefactors would not affect the semiclassical statements in this paper, and they are subleading to quantum corrections that we are ignoring, so we do not keep track of them.}
\begin{equation}
    \boxed{\mathcal{Z}_\text{no-boundary}=2 \int_\mathcal{C} \d N\,\mathcal{G}_N(a_\text{max}\rvert a_\text{max})+2 \int_\mathcal{C}\d N\,\mathcal{G}_N(a_\text{max}\rvert -a_\text{max})\,,}\label{2.20Z}
\end{equation}
As compared to equation \eqref{2.15rho}, we have accounted for sign symmetry in the propagator when $a_0^2=a_1^2$. Integrating over different values of $a_0^2$ at which to glue is a gravitational redundancy as this corresponds to slicing the gravitational solution \eqref{2.17solfinal} at different $\tau$ values.\footnote{Even if one would not view this as redundancy, it would only produce a finite volume prefactor which does not contribute meaningfully to the semiclassical action which we will be investigating.} In section \ref{sect2.2exact}, we shall evaluate the propagators $\mathcal{G}_N(a\rvert \pm a)$. We then take a semiclassical limit of the exact expression to obtain effective actions
\begin{equation}
    \mathcal{G}_N(a\rvert a)\to e^{-I_{a\to a}(N)}\,,\quad \mathcal{G}_N(a\rvert -a)\to e^{-I_{a\to -a}(N)}\,.\label{2.22actionlimit} 
\end{equation}
Later, in section \ref{sect2.4necklasesemi}, we will see that these effective actions $I_{a\to\pm a}(N)$ may be found by solving only the $a$ equations of motion \eqref{2.9aeq} subject to boundary conditions \eqref{2.16bc}, and plugging these into the gravity action \eqref{2.8action}. One can then evaluate the lapse integral in equation \eqref{2.20Z} semiclassically, by performing a Picard-Lefschetz analysis (steepest descent). We will show that Lorentzian contours result in reasonable answers. For the contour $\mathcal{C}=\varepsilon+\i \mathbb{R}$ the saddle $k=1$ dominates $\mathcal{Z}_\text{no-boundary}$, in agreement with CLPW. For the contour $\mathcal{C}=-\varepsilon+\i\mathbb{R}$ the trivial ``saddle'' $k=0$ will dominate. As explained in the introduction section \ref{sect1.2proposal} and expanded upon in the discussion section \ref{sect5:concl}, this would appear a reasonable answer when studying $\rho_\text{no-boundary}(a_0^2\rvert a_1^2)$ for the purposes of understanding the global Hilbert space (which no single observer has access to, but which might nevertheless be of independent interest).

\subsection{Resolution 1: Exact quantum amplitude}\label{sect2.2exact}
We start with an exact evaluation of the propagator $\mathcal{G}_N(a_1\rvert a_2)$:
\begin{equation}
    \mathcal{G}_N(a_0\rvert a_1)=\bra{a_1}e^{-\frac{N}{G} H_\text{WDW}}\ket{a_0}\,.
\end{equation}
Recall the action \eqref{2.8action}. Introduce the Euclidean momentum
\begin{equation}
    p=-\frac{\dot{a}}{N}\,.
\end{equation}
The Euclidean action becomes
\begin{equation}
    I=\frac{1}{G}\int_{0}^{1}\d \tau\big(\, p\, \dot{a}+N H_\text{WDW}\big)\,,\quad H_\text{WDW}=\frac{p^2}{2}+V(a)\,.\label{2.23actionlor}
\end{equation}
$p$ acts on wavefunctions $\psi(a)$ as: $p=-G\,\d/\d a$. One fixes the propagator $\mathcal{G}_N(a_1\rvert a_2)=\bra{a_2} e^{-\frac{N}{G} H_\text{WDW}}\ket{a_1}$ following \cite{halliwell1989multiple,Halliwell:1988wc} by demanding that it satisfies the Schrodinger equation:
\begin{equation} \label{eq:Schr_eq_for_G}
-G \frac{\d}{\d N} \mathcal{G}_N(a_1\rvert a_2) = \frac{1}{2} \bigg( G^2 \frac{\d ^2}{\d a_2^2} -1+a_2^2+\frac{q^2}{4}\frac{1}{a_2^2} \bigg) \,\mathcal{G}_N(a_1\rvert a_2)\,.
\end{equation}
With the boundary condition
\begin{equation}
    \mathcal{G}_{N\to 0}(a_1\rvert a_2)=\left\langle a_2\rvert a_1\right\rangle\,,
\end{equation}
quite non-trivially one obtains the following solution:\footnote{The general differential equation has a one-parameter family of solutions labeled by energy $H_\text{WDW}=E$ eigenvalues. The boundary condition essentially fixes the complete expansion into Hamiltonian eigenfunctions (using delta ortho-normal eigenfunctions $\psi_E(a)$) to have trivial integration kernel:
\begin{equation}
    \mathcal{G}_{N}(a_1\rvert a_2)=\int\d E\,\overline{\psi_E(a_1)}\psi_E(a_2)\,e^{-\frac{N}{G} E}\,.
\end{equation}
This is unique. The Schrodinger equation alone (in both $a_1$ and $a_2$) generally still allows for a free integration kernel $\rho(E)$.
}
\begin{equation} \label{eq:PI_sol}
\mathcal{G}_N(a_1\rvert a_2) = \frac{2 \sqrt{-a_1a_2}}{\sin(N)} \exp\bigg\{ \frac{N}{2G} +\frac{(a_1^2+a_2^2)}{2G\tan(N)} \bigg\} K_{\nu } \bigg( \frac{a_1a_2}{G\sin(N)} \bigg)\,,\quad \nu = \sqrt{\frac{1}{4}-{\frac{q^2}{4G^2}}}\,.
\end{equation}
One checks using the differential equation satisfied by the modified Bessel function $K_\nu(x)$ that, indeed, \eqref{eq:PI_sol} satisfies the Schrodinger equation \eqref{eq:Schr_eq_for_G}. There is another solution $I_\nu(x)$, which was advocated by \cite{halliwell1989multiple} as being the correct propagator. We argue that crucially the correct propagator involves $K_\nu(x)$. Using the Bessel asymptotics and equation \eqref{eq:PI_sol} one indeed obtains for $N\to 0$
\begin{equation}
    \mathcal{G}_N(a_1\rvert a_2)\to 2\pi  \frac{\sqrt{G}}{\sqrt{-2\pi N}} e^{\frac{(a_1-a_2)^2}{2NG} } \to 2\pi G \, \delta (a_1-a_2)\sim \left\langle a_2\rvert a_1\right\rangle\,.\label{2.27delta}
\end{equation}
Instead, $I_\nu(x)$ would result in $\delta(a_1+a_2)$, the wrong result for a quantum mechanical propagator in the limit $N\to 0$. This selects the solution \eqref{eq:PI_sol}.

As discussed in equation \eqref{2.22actionlimit} and equation \eqref{1.13contours1}, we now consider the semiclassical $G\to 0$ limit, to compute an effective action $I_{a\to-a}(N)=-\log \mathcal{G}_N(a\rvert-a)$ and $I_{a\to a}(N)=-\log \mathcal{G}_N(a\rvert a)$. For $G\to 0$ we observe that $\nu\to \i q/2 G$. For this range of parameters, we deduce the following Bessel asymptotics:
\begin{equation}
K_{\i z/G} (x/G) =\frac{1}{2}\int_{-\infty}^{+\infty } \d T \, \exp\bigg\{-\frac{x}{G}\cosh(T)+\i\frac{z}{G}T\bigg\}\to \exp\bigg\{-\frac{x}{G}\sqrt{1-z^2/x^2}-\frac{z}{G}\arcsin(z/x)\bigg\}\,.
\end{equation}
Here we used that the integral representation for $G\to 0$ is dominated by a saddle point: $\sinh(T)=\i z/x$. Inserting this classical approximation in equation \eqref{eq:PI_sol} gives the following effective action:\footnote{We have implicitly assumed the parameters are so that we are describing Euclidean configurations, otherwise the Bessel function asymptotics is governed by two oscillatory saddles. For real $N$ and for $a_1^2=a_2^2=a_\text{max}^2$ this is the correct regime.}
\begin{equation}
    G I_{a_1 \to a_2}(N) = -\frac{N}{2} -\frac{a^2_1+a_2^2}{2}\cot(N) +\frac{a_1 a_2}{\sin (N)} \sqrt{1- \frac{q^2}{4}\frac{\sin^2 (N)}{a_1^2 a_2^2} }+ \frac{q}{2}\arcsin \bigg( \frac{q}{2}\frac{\sin(N)}{a_1 a_2}\bigg)\,.\label{Ia1a2}
\end{equation}
The arcsin$(x)$ is on the principle branch close to the origin $N\to 0$ and continued away from that. In the diagonal case $a_1=a_2=a$, this action reduces to:
\begin{equation}
    G I_{a \to a}(N) = -\frac{N}{2} -\frac{a^2}{\tan(N)} +\frac{a^2}{\sin (N)} \sqrt{1- \frac{q^2}{4}\frac{\sin^2 (N)}{ a^4} }+ \frac{q}{2}\arcsin \bigg( \frac{q}{2}\frac{\sin(N)}{a^2}\bigg)\,.\label{2.32Iaa}
\end{equation}
For the $a_1\to -a_2$ configuration, we simply analytically continue this result from $a_2$ to $-a_2$:
\begin{equation}
    G I_{a_1\to -a_2}(N) = -\frac{N}{2} -\frac{a^2_1 +a_2^2}{2}\cot(N) -\frac{a_1 a_2}{\sin (N)} \sqrt{1- \frac{q^2}{4}\frac{\sin^2 (N)}{a_1^2 a_2^2} }- \frac{q}{2}\arcsin \bigg( \frac{q}{2}\frac{\sin(N)}{a_1 a_2}\bigg)\,.\label{Ia1-a2}
\end{equation}
We are almost exclusively interested in the diagonal case:
\begin{equation}
    \boxed{G I_{a \to -a}(N) = -\frac{N}{2} -\frac{a^2}{\tan(N)} -\frac{a^2}{\sin (N)} \sqrt{1- \frac{q^2}{4}\frac{\sin^2 (N)}{ a^4} }- \frac{q}{2}\arcsin \bigg( \frac{q}{2}\frac{\sin(N)}{a^2}\bigg)\,}\label{2.33Ia-a}
\end{equation}
Note that \eqref{Ia1a2} has square root branch points at $\sin(N) = \pm 2 a_1 a_2/q$, which for $\abs{a_1}=\abs{a_2}=a_\text{max}$ and $0\leq q< 1$ (our case of interest) are off the real $N$ axis. The branch cuts are associated with Stokes lines for the Bessel functions: the function becomes oscillatory. The two sides of the branch cut both contribute (this is obvious because $K_{\i z}(x)$ is real for real $x$). These branch cuts are largely irrelevant for our present purposes, however.

For the purposes of doing a GH calculation, as explained around equation \eqref{2.20Z}, we are especially interested in the scenario where the universe's evolution starts and ends at its maximal Euclidean size:
\begin{figure}[t]
\centering
\includegraphics[width = \textwidth]{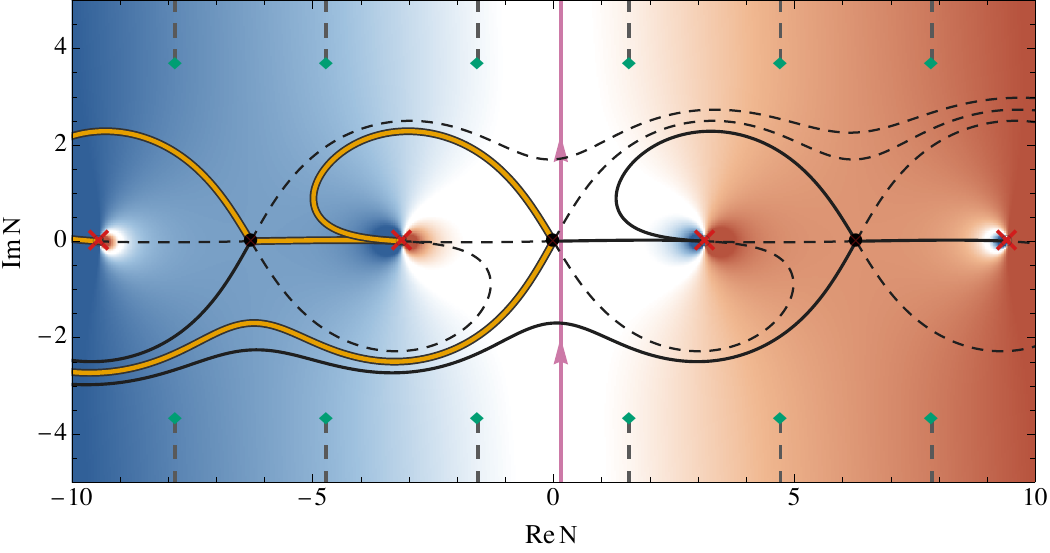}
\caption{Action $I_{a_\text{max} \to a_\text{max}}(N)$ (in \eqref{2.32Iaa}) in the complex $N$ plane. Poles in the action are red $\cross$'s. Saddles are black dots. Steepest descent (ascent) curves are black solid (dashed) lines. The Lorentzian $N=\varepsilon+\i\mathbb{R}$ contour is drawn vertically in magenta and the absolutely convergent contour that it deforms into is shown in orange. Here $G$ was slightly complexified to make the ascent/descent lines non-degenerate.}
\label{fig:plot_plus_sol_N_plane}
\end{figure}
\begin{equation}
    a^2=a_\text{max}^2=\frac{1+\sqrt{1-q^2}}{2}\,.
\end{equation}
The action $I_{a \to a}(N)$ has saddles at $N=2n\pi$ and poles at $N=(2n+1)\pi$ as pictured in equation \eqref{1.14contours2}. The action $I_{a \to -a}(N)$ has saddles at $N=(2n+1)\pi$ and poles at $N=2n\pi$, as shown in equation \eqref{1.13contours1}.  

\subsubsection*{Standard sector}

Figure \ref{fig:plot_plus_sol_N_plane} pictures $I_{a \to a}(N)$ for complex $N$ including saddles and steepest descent lines. In this case, the most natural integration contour is the Lorentzian contour $N=\i \mathbb{R}$. \begin{figure}[h]
\centering
\includegraphics[width=\textwidth]{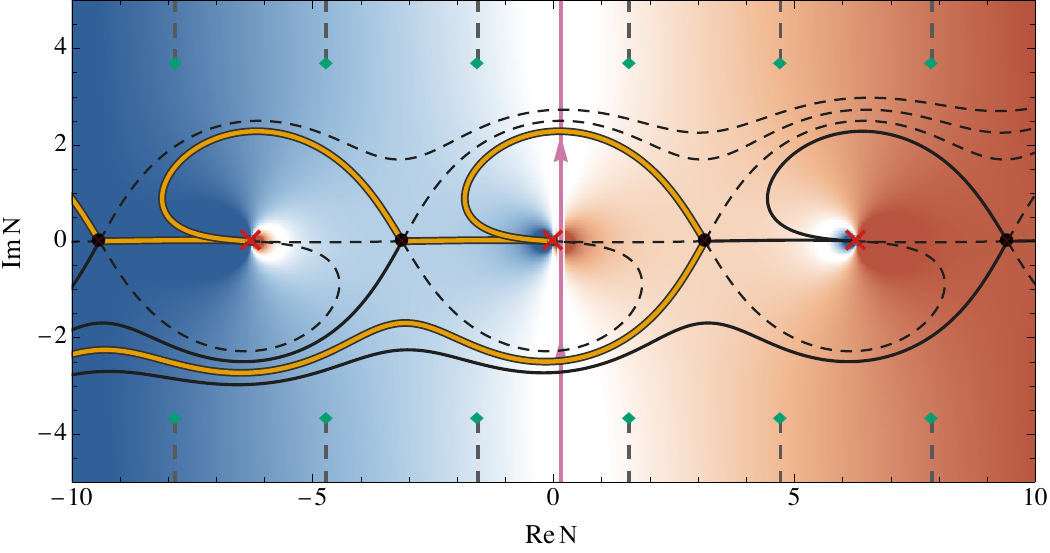}
\caption{Complex $N$ plane for $I_{a_\text{max} \to -a_\text{max}}(N)$ in \eqref{2.33Ia-a}. Singularities in the action are red $\cross$'s and saddles are black dots. Steepest descent (ascent) curves are black solid (dashed) lines. The Lorentzian $N=\varepsilon+\i\mathbb{R}$ contour is drawn vertically in magenta and the absolutely convergent contour that it is deformed into is drawn in orange. Here $G$ was slightly complexified to make the ascent/descent lines non-degenerate. The degenerate real $G$ version is shown in figure \ref{fig:plot_minus_degenerate}.}
\label{fig:plot_minus_sol_N_plane}
\end{figure}
\begin{figure}[h]
\centering
{\includegraphics[width=\textwidth]{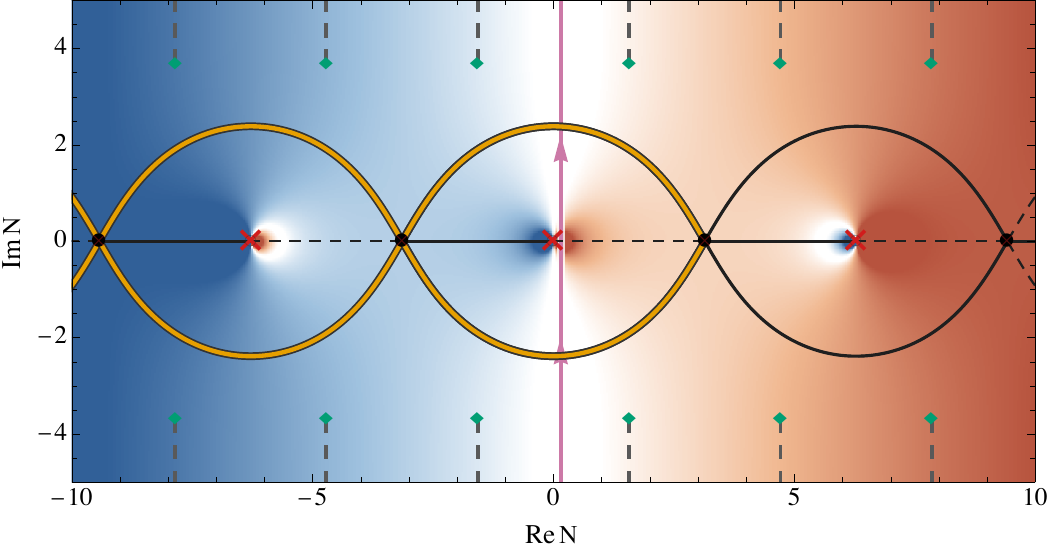}}
\caption{Complex $N$ plane for $I_{a_\text{max} \to -a_\text{max}}(N)$ in \eqref{2.33Ia-a}, with $q=0.1$ and real $G$. A slightly complexified $G$ results in figure \ref{fig:plot_minus_sol_N_plane}. The defining contour deforms to the orange contour dominated by $k=1$, as expected.}
\label{fig:plot_minus_degenerate}
\end{figure}For this contour, the saddle points contribute if $N\leq 0$. The dominant one $N=0$ has $I=0$. The other contributing saddles ($n<0$) have:
\begin{equation}
    I_{a\to a}(2\pi n)=-\frac{\pi n}{G}>0\,.\label{actiona+}
\end{equation}
These contribute negligibly to the path integral. This confirms the picture sketched in equation \eqref{1.14contours2}.
\subsubsection*{Tunneling sector}

The action $I_{a\to -a}(N)$ is shown for complex $N$ in figure \ref{fig:plot_minus_sol_N_plane}. The saddles occur at $N=(2n+1)\pi$. Crucially, there is a singularity at $N=0$, such that there are a priori (at least) two reasonable definitions of the Lorentzian contour: $N=\pm \varepsilon+\i \mathbb{R}$. The question of how the $N$ contour should avoid $N=0$ was recently emphasized in \cite{Banihashemi:2024aal}. For $N=-\varepsilon+\i \mathbb{R}$ one picks up $N<0$ saddles, which are suppressed in the path integral. Crucially, however, for the contour $N=\varepsilon+\i \mathbb{R}$ one picks up the saddle at $N=\pi$, with negative action, and thus positive entropy:
\begin{equation}
    \boxed{I_{a\to-a}(\pi)=-\frac{\pi}{2 G}\overset{?}{=}-S\quad q\text{-independent}\,}\label{action237}
\end{equation}
This is the dominant contribution to the no-boundary amplitude \eqref{2.20Z}, and it gives a positive entropy prediction $S=-I$. This is our main point: a regularized version $N=\varepsilon+\i \mathbb{R}$ of the Lorentzian contour still gives a positive entropy prediction, one that agrees with CLPW \cite{Chandrasekaran:2022cip} when $q\to 0$. Crucially, the pole at $N=0$ avoids the standard argument that negative action saddles could not contribute to the Lorentzian path integral. That argument is that for real configurations, the Lorentzian action is real, and therefore the real part of the Euclidean action vanishes. Because the integration contour could only ``slip down'' to lower values of $\abs{e^{-I}}$, this would usually forbid picking up saddles with $\text{Re}\,I<0$. The $i\varepsilon$ prescription around this $N=0$ singularity avoids this argument. See also section \ref{subsect2.5.4gen}. Note further that $N=+\varepsilon+\i \mathbb{R}$ is the choice of regulator that results in stable matter, in the sense of allowability \cite{Witten:2021nzp,Kontsevich:2021dmb}.

The previous calculation was kind of a ``black box''. We computed the exact propagator and took its semiclassical limit, resulting in effective actions \eqref{2.32Iaa} and \eqref{2.33Ia-a}. But what actually are the spacetime configurations $a(N,\tau)$ responsible for the actions $I_{a\to a}(N)$ and $I_{a\to -a}(N)$? The answer to this question is kind of surprising, and less obvious than what one might expect. To orient the reader, we first discuss the configurations responsible for these actions in the $q\to 0$ dS limit, in section \ref{sec:pure_grav_3d}. We discuss general $q$ configurations in section \ref{sect2.4necklasesemi}.

\subsection{The pure gravity limit}\label{sec:pure_grav_3d}

As claimed around \eqref{2.22actionlimit} we now reproduce the actions $I_{a\to \pm a}(N)$ in equation \eqref{2.32Iaa}/\eqref{2.33Ia-a} by plugging in solutions $a(N,\tau)$ of the $a$ equations of motion \eqref{2.9aeq}, without solving the Friedmann equations (the $N$ equations of motion) \eqref{2.4friedmann}. For empty dS $q\to 0$, the $a$ equations of motion are:
\begin{align}
    \frac{\ddot a}{N^2}+a = 0\,.\label{2.34aeom}
\end{align}
The unique solution $a_+$ with boundary conditions $a_0=a_1=a_\text{max}=1$ (see equation \eqref{2.6max} for $a_\text{max}$) is:
\begin{align}
    a_+=\frac{\cos(N/2-N\tau)}{\cos(N/2)}\,.\label{2.35aplus}
\end{align}
Furthermore, the solution $a_-$ with boundary conditions $a_0=-a_1=a_\text{max}=1$ is:
\begin{align}
    a_-=\frac{\sin(N/2-N\tau)}{\sin(N/2)}\,.\label{2.36amin}
\end{align}
Plugging these into the Einstein-Hilbert action \eqref{2.8action} one obtains:
\begin{align}
    \boxed{GI_{a_\text{max}\to a_\text{max}}(N)=-\frac{N}{2}+\tan(N/2) \,,\quad GI_{a_\text{max}\to -a_\text{max}}(N)=-\frac{N}{2}-\cot(N/2)\,} \label{2.37I}
\end{align}
These reproduce $I_{a\to \pm a}(N)$ in equation \eqref{2.32Iaa}/\eqref{2.33Ia-a} found by taking a semiclassical limit in the exact propagator.

We would now like to understand the meaning of the configurations $a_\pm$ for general $N$. Solutions of the $a$ equations of motion \eqref{2.34aeom} which do not solve the $N$ equations of motion describe the motion of the universe in a quadratic potential:
\begin{equation}
    2V(a)=-1+a^2\,,
\end{equation}
with nonzero energy $E\neq 0$. Indeed, the $N$ equations of motion constrain $E=0$, with $E=p^2/2+V(a)$. Given the trajectory that $a$ covers for $0\leq \tau\leq 1$, we obtain the following convenient relation between the energy $E$ and the elapsed time $N$:
\begin{equation}
    N=\int\frac{\d a}{\sqrt{2 E-2V(a)}}.\label{2.39en}
\end{equation}
One derives this by studying the action \eqref{2.8action} using $(a,p)$ first order variables at fixed energy and writing the symplectic form term as $p\,\d a=\sqrt{2 E-2 V(a)}\,\d a$. Variation with $E$ reproduces the above equation. 

We call the solutions $a_+$ standard solutions and $a_-$ tunneling solutions, for reasons that will become clear in section \ref{sect2.4necklasesemi}.

\subsubsection*{Standard solutions} 

Consider first the configurations $a_+$ in \eqref{2.35aplus}. We discuss the evolution of this universe for $0<N<2\pi$. We will be rather pedantic, as the structure of the configurations $a_\pm$ as a function of $N$ largely survives for $q>0$ (with one important modification regarding the behavior as $a\to0$, see section \ref{sect2.4necklasesemi}). At $N=0$ the universe is static: $a=1$. Otherwise, there are two regimes:
\begin{enumerate}
    \item \underline{$0<N<\pi$} The universe starts at $a=1$. The size of the universe increases until a maximal value $a_\text{largest}=1/\cos(N/2)$. From there, the universe shrinks again until $a=1$. The energy $E$ increases monotonically $\d E/\d N>0$ with $E(0)=0$ and $E(\pi)=\infty$. Configurations with infinite energy are solely responsible for singularities (poles) in the gravitational effective action $I(N)$ for $N\to N_\text{sing}$:
    \begin{equation}
        GI_{a_\text{max}\to \pm a_\text{max}}(N)\to -\frac{2}{N-N_\text{sing}}\to E(N_\text{sing})=+\infty\,. 
    \end{equation}
    This evolution of the universe is described by a particle that goes up the potential $V(a)$ (trajectory ${\color{blue}1}$) before rolling back (trajectory ${\color{blue}2}$):
    \begin{equation}
        \begin{tikzpicture}[baseline={([yshift=-.5ex]current bounding box.center)}, scale=0.7]
    \pgftext{\includegraphics[scale=1]{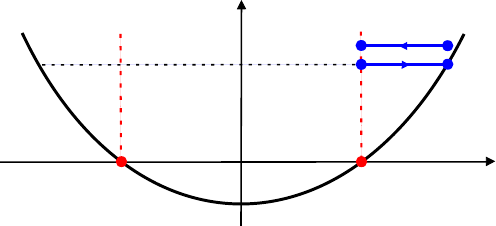}} at (0,0);
    \draw (2.55,1.6) node {\color{blue}$2$};
    \draw (2.55,0.4) node {\color{blue}$1$};
    \draw (2.6,-1.2) node {\color{red}$a_\text{max}$};
    \draw (-2.7,-1.2) node {\color{red}$-a_\text{max}$};
    \draw (4.3,0.4) node {\color{blue}$a_\text{largest}$};
    \draw (-1,1.65) node {$V(a)$};
    \draw (-4.5,0.8) node {$E(N)$};
    \end{tikzpicture} \label{2.41potential}
    \end{equation}
    The Euclidean geometry is one sphere of radius $a_\text{largest}$, cut off at two longitudes where $\abs{a_+}=1$:
    \begin{equation}
        \begin{tikzpicture}[baseline={([yshift=-.5ex]current bounding box.center)}, scale=0.7]
 \pgftext{\includegraphics[scale=1]{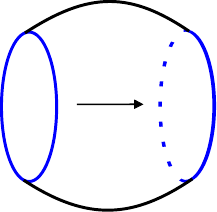}} at (0,0);
    \draw (0,-2.2) node {$0<N<\pi$};
    \draw (-2.7,0) node {\color{blue}$\tau=0$};
    \draw (2.7,0) node {\color{blue}$\tau=1$};
    \draw (2.7,1.5) node {\color{blue}$\abs{a_+}=1$};
    \draw (0,-0.4) node {$\tau$};
    \draw (-1,2) node {$a_+$};
  \end{tikzpicture}
    \end{equation}
    \item \underline{$\pi<N<2\pi$}  The universe starts at $a=1$, and shrinks until $a=0$. Crucially, it then passes into the region $a<0$ (this observation will have important implications for $q>0$ in section \ref{sect2.4necklasesemi}). The size of the universe increases in absolute value, until $a=-\abs{1/\cos(N/2)}$. Then the universe starts shrinking again. It passes through $a=0$, and grows until it reaches $a=1$. This evolution of the universe in the quadratic potential $V(a)$ looks like a particle that makes almost a complete orbit:
    \begin{equation}
        \begin{tikzpicture}[baseline={([yshift=-.5ex]current bounding box.center)}, scale=0.7]
    \pgftext{\includegraphics[scale=1]{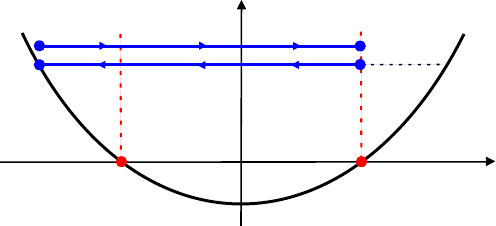}} at (0,0);
    \draw (-1.5,1.6) node {\color{blue}$2$};
    \draw (1,0.4) node {\color{blue}$1$};
    \draw (2.6,-1.2) node {\color{red}$a_\text{max}$};
    \draw (-2.7,-1.2) node {\color{red}$-a_\text{max}$};
    \draw (-4.6,0.4) node {\color{blue}$-a_\text{largest}$};
    \draw (4.4,0.8) node {$E(N)$};
    \draw (-0.1,-2.2) node {$a=0$};\label{2.43pot}
    \end{tikzpicture} 
    \end{equation}
    The Euclidean geometry is three spheres of size $a_\text{largest}$, cut off at two longitudes where $\abs{a_+}=1$:
    \begin{equation}
        \begin{tikzpicture}[baseline={([yshift=-.5ex]current bounding box.center)}, scale=0.7]
 \pgftext{\includegraphics[scale=1]{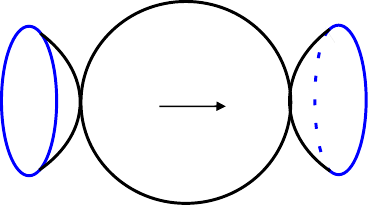}} at (0,0);
    \draw (0,-2.25) node {$\pi<N<2\pi$};
    \draw (-4.1,0) node {\color{blue}$\tau=0$};
    \draw (4,0) node {\color{blue}$\tau=1$};
    \draw (4,1.5) node {\color{blue}$\abs{a_+}=1$};
    \draw (0.2,-0.5) node {$\tau$};
    \draw (-1,2) node {$a_+$};
  \end{tikzpicture}
    \end{equation}
    In this regime, energy decreases monotonically $\d E/\d N<0$, from $E(\pi)=\infty$ until $E(2\pi)=0$. At $N=2\pi$, one finds a gravitational saddle \eqref{2.17solfinal} that describes two copies of the Euclidean sphere.
\end{enumerate}
We stress that the single Euclidean sphere (with action $I=-A/4 G$) does not arise in this $a\to a$ sector. Crucially, the single sphere \emph{does} arise as a solution in the $a\to -a$ sector \eqref{2.36amin}, which we discuss next. 

\subsubsection*{Tunneling solutions} 

Consider the configurations $a_-$ given in equation \eqref{2.36amin}. We describe the evolution of $a_-$ for $0\leq N<2\pi$.
\begin{enumerate}
    \item \underline{$0<N<\pi$} The universe starts at $a=a_\text{max}=1$ and shrinks, until it crunches: $a=0$. Afterwards, the universe is described by $a<0$ and the size grows in absolute value until $a=-a_\text{max}=-1$. In this sector, crucially the solution $N=0$ requires infinite energy: $E(0)=\infty$. Indeed, we force the particle in the quadratic potential (representing the universe's size) to travel a nonzero distance $2a_\text{max}$ in no time. A basic mechanics fact is that this requires an infinite energy. In terms of $V(a)$, the particle travels in a straight line from $a\to -a$:
    \begin{equation}
        \begin{tikzpicture}[baseline={([yshift=-.5ex]current bounding box.center)}, scale=0.7]
    \pgftext{\includegraphics[scale=1]{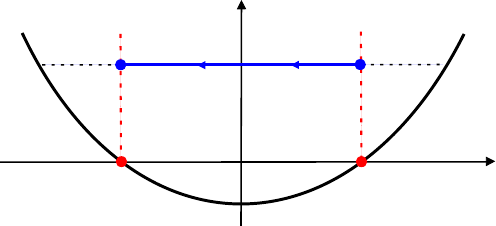}} at (0,0);
    \draw (2.6,-1.2) node {\color{red}$a_\text{max}$};
    \draw (-2.7,-1.2) node {\color{red}$-a_\text{max}$};
    \draw (-1,1.65) node {$V(a)$};
    \draw (-4.5,0.8) node {$E(N)$};\label{2.45pot}
    \end{tikzpicture}
    \end{equation}
    The Euclidean solution consists of two hemispheres touching at the poles, cut at longitudes where $\abs{a_-}=1$:
    \begin{equation}
        \begin{tikzpicture}[baseline={([yshift=-.5ex]current bounding box.center)}, scale=0.7]
 \pgftext{\includegraphics[scale=1]{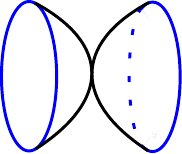}} at (0,0);
    \draw (0,-1.8) node {$0<N<\pi$};
    \draw (-2.4,0) node {\color{blue}$\tau=0$};
    \draw (2.4,0) node {\color{blue}$\tau=1$};
    \draw (2.4,1.5) node {\color{blue}$a_-=-1$};
    \draw (-2.4,1.5) node {\color{blue}$a_-=+1$};
    \draw (0.3,1.3) node {$a_-$};
  \end{tikzpicture}
    \end{equation}
    Energy decreases monotonically $\d E/\d N<0$ until we reach a gravitational saddle $N=\pi$. As we have shown in section \ref{sect2.2exact}, using the contour $N=\varepsilon+\i \mathbb{R}$, this saddle dominates the gravitational path integral and leads to an action consistent with an entropic interpretation according to CLPW:
    \begin{equation}
        I=-\frac{\pi}{2 G}=-\frac{A}{4 G}\,.
    \end{equation}
    \item \underline{$\pi<N<2\pi$} In this sector the universe overshoots. It first grows until $a=1/\sin(N/2)$, then travels all the way to $a=-1/\sin(N/2)$ (passing through $a=0$) before bouncing back to $a=-a_\text{max}=-1$:
    \begin{equation}
        \begin{tikzpicture}[baseline={([yshift=-.5ex]current bounding box.center)}, scale=0.7]
    \pgftext{\includegraphics[scale=1]{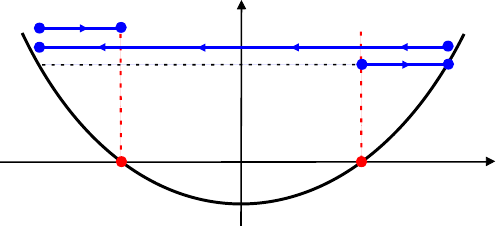}} at (0,0);
    \draw (-3,1.9) node {\color{blue}$3$};
    \draw (0.85,1.6) node {\color{blue}$2$};
    \draw (2.55,0.4) node {\color{blue}$1$};
    \draw (2.6,-1.2) node {\color{red}$a_\text{max}$};
    \draw (-2.7,-1.2) node {\color{red}$-a_\text{max}$};
    \draw (4.3,0.4) node {\color{blue}$a_\text{largest}$};
    \draw (-1,1.65) node {$V(a)$};
    \draw (-4.5,0.8) node {$E(N)$};
    \end{tikzpicture} \label{2.48potential}
    \end{equation}
    The energy grows monotonically in this domain: $\d E/\d N>0$. The singular configuration $N=2\pi$ corresponds to a completely closed orbit in $a$ space, see equation \eqref{2.39en}. One indeed finds that:
    \begin{equation}
        N_\text{orbit}=\oint \frac{\d a}{\sqrt{2 E-2V(a)}}=2\pi\,.
    \end{equation}
    One way to compute this is by contour deformation, picking up the pole at $a\to\infty$. This will be useful in section \ref{subsect2.5.4gen}.
\end{enumerate}

We remark that from the point of view of a particle moving in a quadratic potential $V(a)$, combining equations \eqref{2.41potential}/\eqref{2.43pot}/\eqref{2.45pot}/\eqref{2.48potential}, we conclude that it is certainly \emph{not} true that all trajectories from $a_\text{max}\to\pm a_\text{max}$ contribute in the exact quantum propagator. For instance, take a trajectory in equation \eqref{2.48potential} without an overshoot in the $a<0$ part. Such a trajectory apparently does not contribute in the gravitational path integral. This conclusion persists for $q>0$ (see section \ref{sect2.4necklasesemi}). In 4d and 5d in section \ref{sect4:magnetic} and section \ref{sect3:5daxion}, where we will not have the exact quantum propagator at our disposal, we shall use this output from the 3d analysis as (a proposed) input on which topological sectors of solutions to consider.

This concludes our discussion of the gravitational configurations responsible for the action $I_{a\to \pm a}(N)$ in equation \eqref{2.32Iaa}/\eqref{2.33Ia-a} for empty dS ($q\to 0$). This is part of the process of learning which solutions to count in the gravitational path integral. We now consider $q>0$.

\subsection{Resolution 2: Semiclassical necklace analysis}\label{sect2.4necklasesemi}
The goal of this subsection is to generalize the analysis of section \ref{sec:pure_grav_3d} to $0<q<1$, and identify precisely which configurations are responsible for the effective actions $I_{a\to a}(N)$ and $I_{a\to -a}(N)$ which we obtained from the exact quantum propagators in section \ref{sect2.2exact}. Recall the $a$ equations of motion \eqref{2.9aeq}:
\begin{equation}
    \frac{1}{N^2}\ddot{a}+a-\frac{q^2}{4}\frac{1}{a^3}=0\,.
\end{equation}
The generalization of the $a_+$ dS solution \eqref{2.35aplus} is:
\begin{equation}
    a_+^2=\sqrt{c_+ ^2+\frac{q^2}{4}}+c_+\cos(2N(\tau-1/2))\,,\quad \sin^2(N)c_+ =-a_0^2\cos(N)+ \sqrt{a_0^4-\frac{q^2}{4}\sin^2(N)}\,.\label{2.51aplus}
\end{equation}
The naive generalization of the $a_-$ dS solution \eqref{2.36amin} is:
\begin{equation}
    a_-^2=\sqrt{c_- ^2+\frac{q^2}{4}}+c_-\cos(2N(\tau-1/2))\,,\quad \sin^2(N)c_- =-a_0^2\cos(N)- \sqrt{a_0^4-\frac{q^2}{4}\sin^2(N)}\,.\label{2.52amin}
\end{equation}
Both naively satisfy boundary conditions $a_0=a_1$. Indeed, unlike the dS solution \eqref{2.36amin}, when $0<q<1$ the solution $a_-$ in equation \eqref{2.52amin} has constant sign when $0\leq\tau\leq 1$: 
\begin{equation}
    a_-^2>0\,,\quad 0\leq \tau\leq 1\,.
\end{equation}
Eventually, we will nonetheless identify $a_-$ as representing the configurations with boundary conditions $a_0=-a_1=a$ responsible for $I_{a\to -a}(N)$. This identification will require a modification of the $\tau$ contour. For now, we consider the solutions \eqref{2.51aplus}/\eqref{2.52amin} with $0\leq \tau\leq 1$. Notice that for $q\to 0$, they reduce to their dS counterparts \eqref{2.35aplus}/\eqref{2.36amin}. The action \eqref{2.8action} reduces using the $a$ equation of motion to:
\begin{equation}
    G I_{\pm}=-\frac{N}{2}+N \frac{q^2}{4}\int_0^1\frac{\d\tau}{a_\pm^2}-\frac{1}{2 N}a_\pm\dot{a_\pm}\rvert_{\tau=0}^{\tau=1}\,.\label{2.54I}
\end{equation}

\subsubsection*{Standard solutions} 

Plugging in \eqref{2.51aplus} one obtains (we changed notation $a_0\to a$), for $-\pi<N<\pi$\footnote{Use
\begin{equation}
    \int_{-N}^{+N} \frac{\d \tau}{A+\cos(\tau)} = -\frac{4}{\sqrt{1-A^2}}\text{arctanh}\bigg(\frac{A-1}{\sqrt{1-A^2}}\tan(N/2)\bigg)\,.
\end{equation}
}
\begin{equation}
    G I_+(N) = -\frac{N}{2} -\frac{a^2}{\tan(N)} +\frac{a^2}{\sin (N)} \sqrt{1- \frac{q^2}{4}\frac{\sin^2 (N)}{ a^4} }+ \frac{q}{2}\arcsin \bigg( \frac{q}{2}\frac{\sin(N)}{a^2}\bigg)=G I_{a\to a}(N)\,.
\end{equation}
Comparing with \eqref{2.32Iaa}, this matches the action that follows from the exact propagator with boundary conditions $a_0=a_1=a$. So, $a_+$ is the configuration contributing to this propagator. A classical saddle of this action is $N=2\pi$.

\subsubsection*{Tunneling solutions} 

What about the $a_-$ solution? The on-shell action becomes
\begin{align}
    G I_-(N) &= -\frac{N}{2} -\frac{a^2}{\tan(N)} -\frac{a^2}{\sin (N)} \sqrt{1- \frac{q^2}{4}\frac{\sin^2 (N)}{ a^4} }- \frac{q}{2}\arcsin \bigg( \frac{q}{2}\frac{\sin(N)}{a^2}\bigg)+\frac{\pi q}{2}\nonumber\\
    &=GI_{a\to -a}(N)+\frac{\pi q}{2}\,.\label{2.60}
\end{align}
This action differs significantly from $I_{a\to a}(N)$. So, despite satisfying boundary conditions $a_0=a_1=a$, the solution $a_-$ does \emph{not} contribute to the propagator $\mathcal{G}_N(a\rvert a)$. Instead, we observe that (in analogy with its dS counterpart \eqref{2.36amin}) the solution \emph{almost} reproduces the action $I_{a\to -a}(N)$, corresponding to boundary conditions $a_0=-a_1=a$. There are two issues with such an identification. Firstly, the solution $a_-$ must become negative along $0\leq \tau\leq 1$. Secondly, the actions differ by a constant addition $\pi q/2$. We now solve both issues simultaneously.

Consider the solution $a_-$ along the contour $\tau=1/2+\i T$. This is a crunching Lorentzian spacetime with a Milne-type singularity $a_-=0$ at $T=T_\text{sing}$. It seems logical that to pass from $a_->0$ to $a_-<0$ (in order to respect boundary conditions $a_0=-a_1=a$), one should somehow pass by this singularity. The right procedure is to consider the solution $a_-$ in equation \eqref{2.52amin} along a $\tau$ contour that winds once around the big-crunch singularity:
\begin{equation}
    \begin{tikzpicture}[baseline={([yshift=-.5ex]current bounding box.center)}, scale=0.7]
 \pgftext{\includegraphics[scale=1]{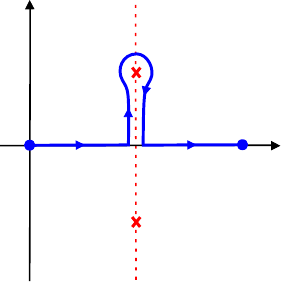}} at (0,0);
    \draw (1.7,1.2) node {\color{red}big crunch};
    \draw (-1.1,1.2) node {\color{red}$T_\text{sing}$};
    \draw (2.2,-0.5) node {$\tau$};
    \draw (-2.2,-0.5) node {\color{blue}$0$};
    \draw (1.5,-0.5) node {\color{blue}$1$};
    \draw (1.3,-1.35) node {\color{red}big bang};
  \end{tikzpicture}\label{2.59contour}
\end{equation}
An instructive example of this configuration is the classical saddle of the action $I_-(N)$ (at $N=\pi$) with boundary conditions $a^2=a_\text{max}^2$. Solving for $c_-$ in equation \eqref{2.52amin} one recovers the classical solution \eqref{2.5Euclideansol}:
\begin{equation}
    a_-^2=\frac{1}{2}+\frac{\sqrt{1-q^2}}{2}\cos(2\pi\tau)\,.
\end{equation}
Along the contour $\tau=1/2+\i T$ one now finds the following big-bang big-crunch Lorentzian spacetime:
\begin{equation}
     a_-^2=\frac{1}{2}-\frac{\sqrt{1-q^2}}{2}\cosh(2\pi T)\qquad \begin{tikzpicture}[baseline={([yshift=-.5ex]current bounding box.center)}, scale=0.7]
 \pgftext{\includegraphics[scale=1]{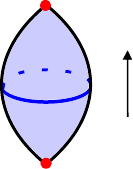}} at (0,0);
    \draw (1.3,1.5) node {\color{red}big crunch};
    \draw (1.6,0) node {$T$};
    \draw (1.23,-1.5) node {\color{red}big bang};
  \end{tikzpicture}
\end{equation}
The location of the singularities is
\begin{equation}
    T_\text{sing}=\frac{1}{2\pi}\text{arccosh}\left(\frac{1}{\sqrt{1-q^2}}\right)\,.
\end{equation}
Crucially, expanding around $T\to T_\text{sing}$ gives $a_-(T) \sim (T-T_\text{sing})^{1/2}$. Hence, we see that, going around the singularity once as instructed by the $\tau$ contour in equation \eqref{2.59contour}, $a_-$ acquires an additional phase: $a_-\to e^{\i \pi} a_-$. This proves that along the contour \eqref{2.59contour}, $a_-$ is indeed a solution with boundary conditions $a_0=-a_1=a$.

Crucially, the on-shell gravitational action $I_-$ \eqref{2.54I} receives a nontrivial extra contribution due to winding around the singularity. Using equation \eqref{2.52amin}, somewhat miraculously one finds:
\begin{equation}
    GI_-(N)\to G I_-(N)+ N\frac{q^2}{4}\oint_{a_-=0}\frac{\d \tau}{a_-^2} =G I_-(N)-\frac{\pi q}{2}=GI_{a\to -a}(N)\,.\label{corrrr}
\end{equation}
This is the main result of this section: the solution $a_-$ \eqref{2.52amin} along a bra-ket wormhole contour \eqref{2.59contour} satisfies boundary conditions $a_0=-a_1=a$ and is responsible for the exact propagator $\mathcal{G}_N(a\rvert -a)$. An important point was that there was a contribution from winding around the singularity in the complex $a$ plane. As explained in section \ref{sect2.2exact}, a steepest descent analysis now reveals that the dominant contribution to the gravitational path integral is the solution $a_-$ (with $N=\pi$).

We stress that this Lorentzian detour, and the associated correction \eqref{corrrr} to the action, is essential to obtain a $q$-independent on-shell action \eqref{action237}. We discuss the puzzling $q$-independence of the entropy in more detail in section \ref{sect:3.3}. The $a\to a$ sector with action $I_{a\to a}(N)$ should be interpreted in a similar manner: like the empty dS solution \eqref{2.43pot}, the solutions with $N=2\pi n$ crosses $a=0$ a total $2n$ times. This gives the $q$-independent necklace action \eqref{actiona+}.

We remark that the action is unique provided $I_{a_1\to a_2}(N\in \i \mathbb{R})\in \i\mathbb{R}$ for configurations $a_1,a_2$ associated with Lorentzian solutions. In particular, one could not simply add terms like $\pi q/2$ to the action for all $N$. This makes for instance equation \eqref{2.33Ia-a} robust, and thus also our key (but confusing) result \eqref{key} below.

\subsubsection*{Dominant solution}

This dominant solution looks as follows:
\begin{equation}
    \begin{tikzpicture}[baseline={([yshift=-.5ex]current bounding box.center)}, scale=0.7]
    \pgftext{\includegraphics[scale=1]{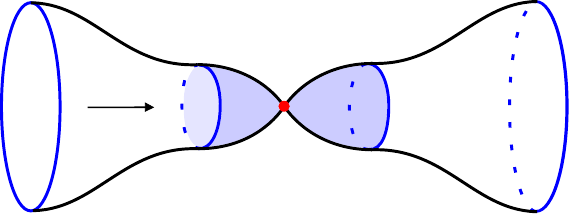}} at (0,0);
    \draw (-9,0) node {$\mathcal{Z}_\text{no-boundary}\to$};
    \draw (0,-1.2) node {\color{blue}Lorentzian};
    \draw (-2.6,-0.4) node {$\tau$};
    \draw (5.6,0) node {\color{blue}glue};
    \draw (4.7,2.1) node {\color{blue}$a_-=-a_\text{max}$};
    \draw (-3.9,2.1) node {\color{blue}$a_-=+a_\text{max}$};
    \draw (2.2,1.3) node {$a_-$};
    \draw (-5.6,0) node {\color{blue}glue};
    \end{tikzpicture} 
\end{equation} 
In the dS limit the Lorentzian region shrinks away to a point, resulting in the usual sphere amplitude. If the no-boundary path integral computes an entropy, as suggested by \cite{witten2024background,Blommaert:2025bgd}, that entropy would be the on-shell action associated with this geometry:
\begin{equation}
    \boxed{S=\frac{\pi}{2 G}\,}\label{key}
\end{equation}
This is a clear target for Lorentzian algebraic calculations.

We started out in section \ref{sect2.2exact} with describing the universe as a quantum mechanical particle traveling in a potential $V(a)$ \eqref{potential}:
\begin{equation}
    2V(a)=-1+a^2+\frac{q^2}{4}\frac{1}{a^2}\,.
\end{equation}
The generalization of the dS motion of the particle in the potential, discussed in equation \eqref{2.45pot}, is the following motion:
\begin{equation}
    \begin{tikzpicture}[baseline={([yshift=-.5ex]current bounding box.center)}, scale=0.7]
    \pgftext{\includegraphics[scale=1]{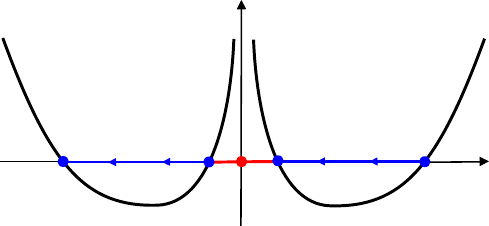}} at (0,0);
    \draw (3.7,-1.3) node {\color{blue}$a_\text{max}$};
    \draw (1.2,-0.35) node {\color{blue}$a_\text{min}$};
    \draw (0.8,1.65) node {$V(a)$};
    \draw (-0.4,-1.25) node {\color{red}$0$};
    \end{tikzpicture}\label{2.69tunnel}
\end{equation}
From this perspective, the Lorentzian bra-ket wormhole piece of the geometries is viewed as a {\color{red}tunneling} solution. Because we started from Euclidean signature, the tunneling process is in Lorentzian signature, the opposite of what happens usually. Other off-shell values of $N$ correspond to configurations $a_-$ in equation \eqref{2.52amin} which describe the motion of the particle higher up in the potential, analogous to what happened in empty dS, see equation \eqref{2.45pot}.
The $q\to 0$ pure dS limit is similar to the operator insertion limit of \cite{Witten:2026twr} (see also \cite{Ivo:2026ijv}).

We briefly reiterate the logic. We studied the exact quantum propagator, and used this to identify which classical configurations $a_+$ and $a_-$ contribute to respectively the propagators $\mathcal{G}_N(a\rvert a)$ and $\mathcal{G}_N(a\rvert -a)$. The remainder of this work concerns generalizations where we do not have access to exact quantum propagators, but where instead we assume these same ``topological sectors'' of solutions are responsible for $\mathcal{G}_N(a\rvert a)$ and $\mathcal{G}_N(a\rvert -a)$.

\subsection{General argument}\label{subsect2.5.4gen}
Consider the regime $N\to 0$ for the Euclidean action \eqref{2.8action} describing propagation in a generic potential:
\begin{equation}
    I=\frac{1}{\text{G}}\int_0^1\d \tau \bigg\{ -\frac{1}{2N}\dot{a}^2+N V(a) \bigg\}\,.
\end{equation}
The $N\to 0$ approximation is to drop the potential term when computing $\mathcal{G}_N(a_1\rvert a_2)$, as a particle needs very high kinetic energy to travel from $a_1\to a_2$ in almost zero time $N$. The solution is $a=a_1+(a_2-a_1) \tau$, leading to the leading order $N\to 0$ approximation generalizing equation \eqref{2.27delta}:
\begin{equation}
\boxed{\mathcal{G}_N(a_1\rvert a_2)\,\to\, e^{\frac{1}{2 N \text{G}}(a_1-a_2)^2}\,}\label{2.77pole}
\end{equation}
For $\mathcal{G}_N(a\rvert a)$ the action vanishes and indeed we did not observe a singularity at $N=0$, rather the action develops a classical saddle (the particle not moving). Crucially, for $\mathcal{G}_N(a\rvert -a)$, this action does develop an $N\to 0$ singularity, as shown in equation \eqref{2.77pole}.

More generally, consider any FLRW potential of the form \eqref{potential}, with quadratic asymptotics. Two ingredients survive from the 3d analysis. First, the zero-energy bounce has some period $N_0$, and gluing $k$ bounces produces necklace saddles at $N=kN_0$, with odd $k$ describing the $a\to-a$ sector and even $k$ the $a\to a$ sector. The former can be viewed as tunneling through $a=0$. Secondly, the actions $I_{a\to\pm a}(N)$ are real on the real $N$ axis and the $N=0$ pole \eqref{2.77pole} is there in the $a\to -a$ sector only. The segment $(0,N_0)$ of the real axis is a steepest ascent line of the $k=1$ saddle. This crosses the contour $N=\varepsilon+\i\mathbb{R}$. Picard-Lefschetz therefore guarantees that it contributes to the path integral.\footnote{At $a=a_\text{max}$ the saddle is cubic. It may be convenient to view this as a limit of Lorentzian boundary conditions $a>a_\text{max}$ in which case there are two saddles on opposite sides of the real axis, which coincide when $a\to a_\text{max}$. See figure \ref{fig:app2}. For Euclidean $a<a_\text{max}$ boundary conditions, two saddles are on the real axis, and according to the Picard-Lefschetz argument, only the first saddle contributes. This transition matches the Stokes phenomenon in the asymptotics of the Bessel function \eqref{eq:PI_sol}.}

\subsubsection*{Walls}

The more difficult part is proving that the higher necklaces ($k\geq 2$) will not contribute. In other words, ascent lines from $k\geq 2$ may not cross $N=\varepsilon+\i\mathbb{R}$. For the analytically solvable axion model one can understand this analytically. Consider the actions \eqref{2.32Iaa}/\eqref{2.33Ia-a}. We focus on $I_{a_\text{max}\to -a_\text{max}}(N)$. Define the \textbf{height function}
\begin{equation}
    H(N)=-\text{Re}\,I_{a_\text{max}\to -a_\text{max}}(N)\,.
\end{equation}
Ascent lines $\mathcal{A}_k$ follow lines of constant $\text{Im}\,I_{a_\text{max}\to -a_\text{max}}(N)$, start at $N=k N_0$, and have monotonically increasing height functions (they are ascending). Therefore (recalling that saddles have odd $k=2m+1$):
\begin{equation}
    H(\mathcal{A}_{2m+1})>(2m+1) \abs{I_0}\,.\label{2.74confine}
\end{equation}
The function $I_{a_\text{max}\to -a_\text{max}}(N)$ given in equation \eqref{2.33Ia-a} has a remarkable property. The height function $H(N)$ is constant on lines above and below the singularities $N=2m N_0$:
\begin{equation}
    H(2m N_0+\i \mathbb{R})=2m \abs{I_0}\,.\label{walls}
\end{equation}
Combined with equation \eqref{2.74confine}, we see that the ascent line of the saddle at $N=(2m+1)N_0$ is confined to the region $\text{Re}\, N>2m N_0$:
\begin{equation}
    \text{Re}\,\mathcal{A}_{2m+1}>2 m N_0\,.
\end{equation}
Therefore the ascent line does not intersect the defining contour $\mathcal{A}_{2m+1} \cap \varepsilon+\i \mathbb{R}=\varnothing$. This demonstrates analytically that the higher necklaces $k\geq 2$ indeed do not contribute.

\subsection{Area law universality}\label{sect:3.3}
In equation \eqref{action237}, we found that the dominant saddle $k=1$ gives an entropy independent of the axion flux parameter $q$ equal to the empty dS entropy \eqref{action237}. This might be interpreted as one version of the cosmic no-hair theorem, as we explain next.

The contour used in \eqref{2.59contour} to evaluate the on-shell action may be deformed in the following manner:
\begin{equation}
    \begin{tikzpicture}[baseline={([yshift=-.5ex]current bounding box.center)}, scale=0.7]
 \pgftext{\includegraphics[scale=1]{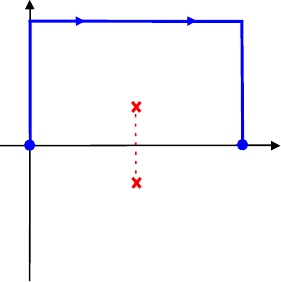}} at (0,0);
    \draw (0,2.5) node {\color{blue}$T\to\infty $};
    \draw (0,3.2) node {\color{blue}pure dS solution};
    \draw (2.2,-0.5) node {$\tau$};
    \draw (-2.2,-0.5) node {\color{blue}$0$};
    \draw (1.5,-0.5) node {\color{blue}$1$};
    \draw (0,-1.25) node {\color{red}big bang};
  \end{tikzpicture}\label{eq:new_time_contour_fig}
\end{equation}
This contour is labeled by a large Lorentzian time $T\to \infty$.
We study the contribution from the $N=\pi$ GH saddle along this contour in the case $a_1 = -a_2 = a_\text{max}$ \eqref{2.6max}. The solution for complex $\tau$ reads:
\begin{equation}
    a^2(\tau) = \frac{1}{2} + \frac{\sqrt{1-q^2}}{2} \cos(2\pi\tau).
\end{equation}
The two Lorentzian contributions $\tau=0\to \i T$ and $\tau=1+\i T\to 1$ exactly cancel. What remains is the late time contour $\tau = \i T + \theta/\pi$
\begin{equation}\label{eq:a_theta}
    a(\theta)\to a_T\,e^{-\i \theta}\,,\quad 0\leq \theta\leq \pi\,,\quad a_T=\frac{(1-q^2)^{1/4}}{2} e^{\pi T}\,.
\end{equation}
Shifting $T\to T+c$ this becomes the $q$-independent pure dS solution, as demanded by the cosmic no-hair theorem \cite{Wald:1983ky,Gibbons:1977mu,Starobinsky:1983zz}. Since $T\to T+c$ is an analytic contour deformation, the value of the action does not change. $T\to T+c$ absorbs the $q$-dependence in $a_T$, so the on-shell action is $q$-independent. Explicitly, substituting \eqref{eq:a_theta} into the $d=3$ action \eqref{2.8action}:
\begin{equation}
\begin{split}
    G I_{a\to -a}(2\pi) &= \int_0^\pi d\theta \left\{-\frac{a_\theta^2}{2} + \frac{1}{2}\left(-1+a^2+\frac{q^2}{4 a^2}\right)\right\}\\
    &= \int_0^\pi d\theta \frac{(1-q^2)\sinh^2(2\pi T - 2i \theta)}{2+2\sqrt{1-q^2} \cosh(2\pi T-2i \theta)}\\
    & \to  \int_0^\pi d\theta \  \left\{a_T^2 e^{- 2i \theta} -\frac{1}{2}\right\}= -\frac{\pi}{2}.
\end{split}
\end{equation}
The integral on the second line is exact and gives $\pi/2$. The one on the last line takes the large $T$ limit and follows from the dS potential $q=0$.

We find the independence of this result on $q$ to be puzzling. However, at least in 3d, this conclusion appears difficult to avoid, despite attempts, and so we may need to reckon with the physical implications. 

\section{Higher dimensional axion necklaces}\label{sect3:5daxion}
In this section, we will provide further evidence for the generality of our claim. We investigate higher-dimensional axion wormholes numerically. Within minisuperspace, we show that the Lorentzian lapse contour $N=\varepsilon+\i \mathbb{R}$ picks up the GH saddle as the dominating contribution.

In \textbf{section \ref{sect3.2kinetic}} we set up the minisuperspace necklace problem in (what we will call) \textbf{kinetic gauge}. In this gauge, the scale factor has a standard kinetic term, and the action possesses $a\mapsto -a$ symmetry. As in the 3d analysis, the quantum gravity propagator follows from the $a(\tau)$ path integral upon gauging this symmetry. Unlike in 3d (section \ref{sect2.2exact}) we do not know the exact analytic propagator. Instead, we rely on the semiclassical lessons of section \ref{sect2.4necklasesemi} as a guide for which trajectories $a(\tau)$ to investigate. The associated action $I(N)$ is studied numerically in 4d and 5d, in \textbf{section \ref{sect3.3kineticnum}}. In appendix \ref{app:5dproper}, we present an alternative analytic and numerical analysis in (what we call) proper-time gauge for 5d.

\subsection{Setup}\label{sect3.2kinetic}
Following section \ref{sect2.1setup}, we consider an axion flux coupled to $\Lambda>0$ gravity. The Euclidean action reads:
\begin{equation}\label{eq:axion_action}
    I = \int \d^d x \sqrt{g} \bigg\{-\frac{1}{16\pi G}\left(R-2\Lambda\right) + \frac{1}{(d-1)!} H^{ij\cdots}H_{ij\cdots}\bigg\}\,.
\end{equation}
Here $H_{ij\cdots} = Q \cdot \text{vol}$ is a $(d-1)$-form field strength and $\text{vol}$ is the $(d-1)$-sphere volume form. For the metric, we will use minisuperspace FLRW with the non-standard \textbf{kinetic gauge}:
\begin{equation}
    \d s^2=N^2 a(\tau)^{2(d-3)}\d \tau^2+a(\tau)^2\d \Omega_{d-1}^2\,,\quad 0\leq \tau \leq 1\,.
\end{equation}
The purpose of this gauge is that the Euclidean action \eqref{eq:axion_action} (with $H=1$) has an ordinary kinetic term.
Indeed, following the conventions of \cite{Aguilar-Gutierrez:2023ril}:
\begin{equation}\label{eq:axion_conv}
    \Lambda = \frac{(d-1)(d-2)}{2}H^2\,,\quad Q^2 = \frac{d-2}{16\pi G} \left( \frac{d-2}{d-1} H^{-2}\right)^{d-2} q^2\,,\quad 0\le q \le 1\,,
\end{equation}
the Euclidean action has a standard $\dot{a}^2$ kinetic term:
\begin{equation}
    \boxed{I = \frac{\Omega_{d-1}(d-1)(d-2)}{8\pi G}\int_0^1 \d\tau\, \bigg\{-\frac{\dot a^2}{2N} +\frac{N}{2}\bigg(-a^{2d-6}+a^{2d-4} + \frac{1}{d-1}\left(\frac{d-2}{d-1}\right)^{d-2} \frac{q^2}{a^2} \bigg)\bigg\}\,}\label{3.10kinetic}
\end{equation}

This action shares key properties with our $d=3$ case. First, since the kinetic term is standard, we can set up a similar well-posed quantum mechanical system, which will reproduce a sensible propagator $\mathcal{G}_N(a_1|a_2)$ for $a(\tau)$. 
Secondly, $a\mapsto -a$ is again a symmetry of the action, so we can gauge it, leading to the same two sectors as in 3d. Thirdly, due to the last term in \eqref{3.10kinetic}, the big-bang solution behaves as $a(t) \sim t^{1/2}$ for any $d$. For this reason, we expect that a contour similar to \eqref{2.59contour} will again map positive to negative $a$. Therefore, the saddles that contribute to each sector follow the same structure as in 3d.

In any dimension, the classical equations of \eqref{3.10kinetic} describe a particle in a potential. Below we study this action numerically for 4d and 5d where the potential equals:
\begin{equation}
    \begin{split}
        2V_\text{4d}(a) &= -a^2+a^4+\frac{4}{27} \frac{q^2}{a^2},\\
        2V_\text{5d}(a) &= -a^4+a^6+\frac{27}{256}\frac{q^2}{a^2}.
    \end{split}
\end{equation}
In both cases, this potential allows for periodic solutions with zero energy. These solutions are exactly the higher-dimensional analogs of the necklaces we found in the previous section.
Working for general lapse $N$/energy, just as in 3d, we can find semiclassical solutions for $a(\tau)$. In the $a\to a$ sector (analogous to equation \eqref{2.41potential} for 3d) the solution would be:
\begin{equation}
        \begin{tikzpicture}[baseline={([yshift=-.5ex]current bounding box.center)}, scale=0.7]
    \pgftext{\includegraphics[scale=1]{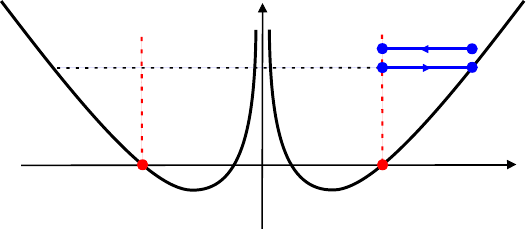}} at (0,0);
    \draw (2.55,1.6) node {\color{blue}$2$};
    \draw (2.55,0.4) node {\color{blue}$1$};
    \draw (2.6,-1.2) node {\color{red}$a_\text{max}$};
    \draw (-2.7,-1.2) node {\color{red}$-a_\text{max}$};
    \draw (4.3,0.4) node {\color{blue}$a_\text{largest}$};
    \draw (-1,1.65) node {$V(a)$};
    \draw (-4.5,0.8) node {$E(N)$};
    \end{tikzpicture} 
    \end{equation}
The tunneling solution $a\to -a$ similarly generalizes the 3d tunneling solution shown in equation \eqref{2.69tunnel}:
\begin{equation}
        \begin{tikzpicture}[baseline={([yshift=-.5ex]current bounding box.center)}, scale=0.7]
    \pgftext{\includegraphics[scale=1]{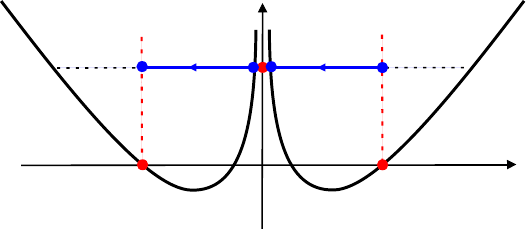}} at (0,0);
    \draw (-1.5,0.4) node {\color{blue}$2$};
    \draw (1.5,0.4) node {\color{blue}$1$};
    \draw (2.6,-1.2) node {\color{red}$a_\text{max}$};
    \draw (-2.7,-1.2) node {\color{red}$-a_\text{max}$};
    \draw (-1,1.65) node {$V(a)$};
    \draw (-4.5,0.8) node {$E(N)$};
    \end{tikzpicture} 
    \end{equation}
In the next section \ref{sect3.3kineticnum} we present the numerical evaluation of the semiclassical actions $I_{a_\text{max}\to \pm a_\text{max}}(N)$. Picard-Lefschetz reveals that a Lorentzian lapse contour $N=\varepsilon+\i \mathbb{R}$ will pick up the GH saddle as the dominating contribution. The $q=0$ dS case may be studied partially analytically. See equation \eqref{4.15sol}.

\subsubsection*{On-shell action}

Our approach in this section is to assume that the solutions are given by the ordinary necklace solutions, as can also be obtained without a minisuperspace approximation, plus the tunneling piece (or several of these for higher necklaces). In the tunneling piece, the Lorentzian parts cancel and we are left with the contribution around $a=0$. In kinetic gauge, this is a pole in any dimension and we can calculate its contribution. Indeed, around $a \approx 0$, only the $1/a^2$ term in the potential is important, leading to a solution $a(\tau) = (-4N^2 q^2 x)^{1/4}\sqrt{\tau}$ where we denote the coefficient in the action $x=\frac{1}{d-1}\left(\frac{d-2}{d-1}\right)^{d-2}$. Therefore for small $a$ (meaning also small $\tau$), plugging back into the action, we see that the action behaves as $ \frac{\Omega_{d-1}(d-1)(d-2)}{8\pi G} \int d\tau \frac{iq\sqrt{x}}{2\tau}$, and so the pole contribution to the entropy is
\begin{equation} \label{eq:pole_contribution_general_d}
    S_{\text{pole}}=\frac{\Omega_{d-1}(d-1)(d-2)}{8\pi G} \pi q \, \sqrt{\frac{1}{d-1}\left(\frac{d-2}{d-1}\right)^{d-2}} .
\end{equation}
For $d=3$ this is indeed $\frac{\pi q}{2G}$.

On the other hand, for pure de Sitter in Euclidean signature, the on-shell action for a single sphere is just
\begin{equation} \label{eq:sphere_action_general_d}
    I_{\text{sphere}} = -\frac{1}{4\pi G(d-2)} \int d^dx \sqrt{g} \, \Lambda = -\frac{\Omega_d (d-1)}{8\pi G}
\end{equation}
where in the second equality we used $H=1$ units.

The Einstein static universe case is the value of $q$ for which there is a minimum of the potential at $V=0$, and this happens at any dimension at $q=1$ in these conventions (and the size of the universe is $a_0=\sqrt{\frac{d-2}{d-1}}$ in this case). At this value, the Euclidean solution has vanishing action (since it is stationary and the potential is zero). Therefore the only contribution of the tunneling solution is from the pole \eqref{eq:pole_contribution_general_d}. We notice that at $d=3$ this value is equal to the entropy of the sphere $S=-I$ from \eqref{eq:sphere_action_general_d} (for $q=0$ there is no contribution from the $a=0$ pole), but at $d>3$ we have $S_{q=1}=S_{\text{pole}}(q=1) > S_{\text{sphere}}$ which would contradict the second law of thermodynamics.

We emphasize that only at $d=3$ we have really shown that the tunneling solution is what governs the quantum mechanics. It may well be that this is not the case for generic $d$. In particular, the pole contribution is linear in $q$ for any $d$, whereas the necklace action is linear in $d=3$ but not for general $d$, which would mean that the total action of the tunneling solution is $q$ independent only for $d=3$ but not for higher $d$. If the tunneling solution is not the right answer in higher dimensions, that would invalidate this conclusion. It would also avoid the violation of the second law discussed above. At $d=3$ where we know the tunneling solution is correct, there is no such violation. See section \ref{sect:3.3} for more discussion.

\subsection{Numerical analysis}\label{sect3.3kineticnum}

\begin{figure}
\centering
\includegraphics[width=.48\textwidth]{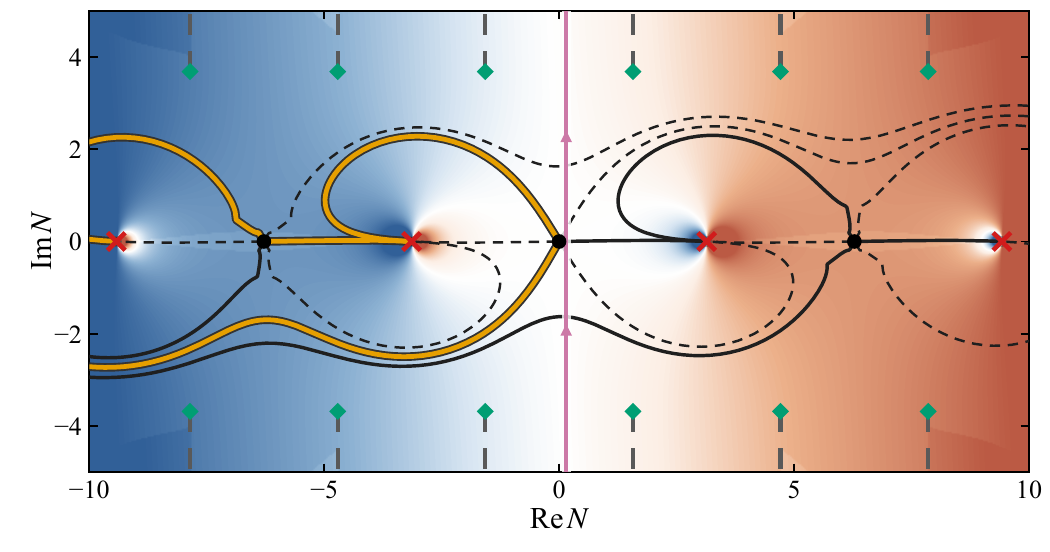}
\includegraphics[width=.48\textwidth]{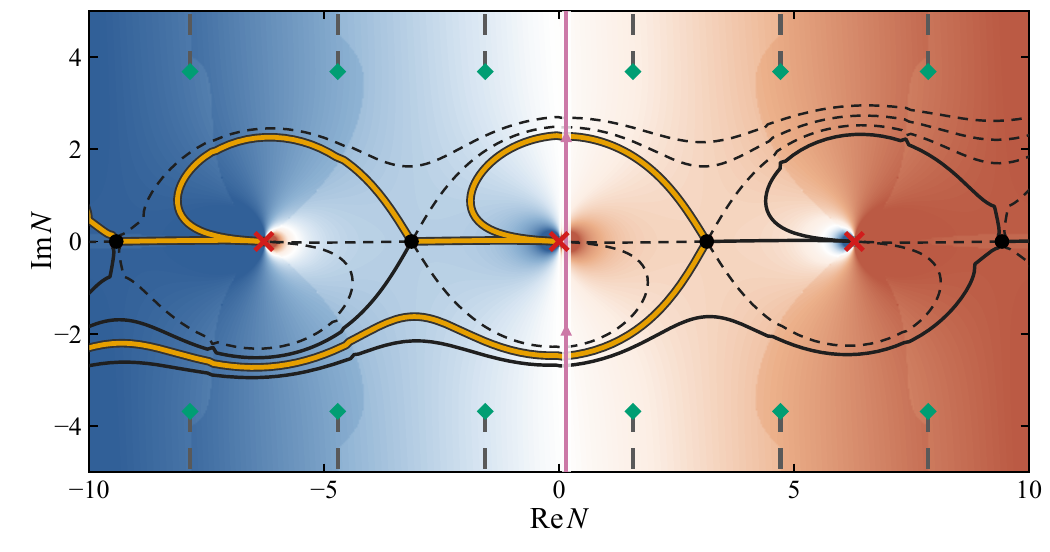}
\caption{Numerical simulations of the 3d action $I_{a_\text{max}\to \pm a_\text{max}}(N)$ show excellent agreement with plots of the analytically obtained on-shell actions (figure \ref{fig:plot_plus_sol_N_plane} and figure \ref{fig:plot_minus_sol_N_plane}). This supports higher dimensional numerics. Here $G$ was slightly complexified to make the ascent/descent lines non-degenerate.}
\label{fig:np33check3d}
\end{figure}

To implement numerically, we solve the $a$ equations of motion as a boundary value problem for general complex lapse $N$ using the shooting method. The action is then evaluated on this solution for each $N$.

Before tackling 4d/5d, we test the procedure in 3d, where we can check against the analytic solution of section \ref{sect2:3daxion}. In figure \ref{fig:np33check3d}, the $N$ plane for the $a_-$ solution is shown, to be compared with figure \ref{fig:plot_minus_sol_N_plane}.\footnote{Technically, we are solving the problem for $a_\text{max}\to a_\text{max}$ boundary conditions and are indeed investigating the so-called $a_-$ bouncing solution. As discussed in section \ref{sect2:3daxion}, around equation \eqref{2.60}, this is related to the tunneling $a_\text{max}\to -a_\text{max}$ solution by the addition of a constant in the action. This structure persists in higher dimensions, because of the universal $a\to 0$ behavior of the potential in kinetic gauge \eqref{3.10kinetic}, giving a universal solution for $a\to 0$ and ditto pole contributions.} The singularity for $N\to 0$ is clear. There is a saddle point on the positive real $N$ axis with steepest ascent line crossing the defining integration contour. More generally, one finds good agreement with analytics, though there are some numerical artifacts. The most significant is that the numerical analysis, in some places, picks up the wrong solution. This issue persists in 4d/5d. However, fortunately, the branch cuts and associated choices of solutions only cloud things (relatively) far away from the origin $N\to 0$.

Consider the height function $H(N)$ defined in section \ref{subsect2.5.4gen}. There are several solutions for fixed $N$, so the action $I(N)$ has many sheets and many branch cuts. This complicates following either the lines of constant height $H(N)= 2m \abs{I_0}$ or the ascent lines $\mathcal{A}_{2m+1}$. We found that one can numerically track individual ascent lines $\mathcal{A}_{2m+1}$, even as they cross sheets, and check whether these eventually cross $N= \varepsilon+\i \mathbb{R}$. We did not find any higher necklaces that contribute. However, a case-by-case numerical investigation of all ascent lines $\mathcal{A}_{2m+1}$ from all saddles is neither practical nor fully airtight.

Instead, one proceeds as follows. On the defining contour $N=\varepsilon+\i \beta$ we have (away from $N\to 0$) $H(\varepsilon+\i \beta)\sim \varepsilon$, as the action is purely imaginary for purely imaginary $N$ (indeed, Lorentzian solutions), and that $I(N)$ has a Taylor expansion (away from $N\to 0$). Including the pole $I(N\to 0)\sim 1/N$, we have (modulo a constant) $H(\varepsilon+\i\beta)\to\delta(\beta)$. Furthermore, notice \eqref{2.74confine} $ H(\mathcal{A}_{2m+1})>(2m+1) \abs{I_0}$, an order one number (in units of $1/G$). So, $\mathcal{A}_{2m+1}$ could only cross the contour $N=\varepsilon+\i \mathbb{R}$ at $N=0$.

\begin{figure}[h]
\centering
\includegraphics[width=\textwidth]{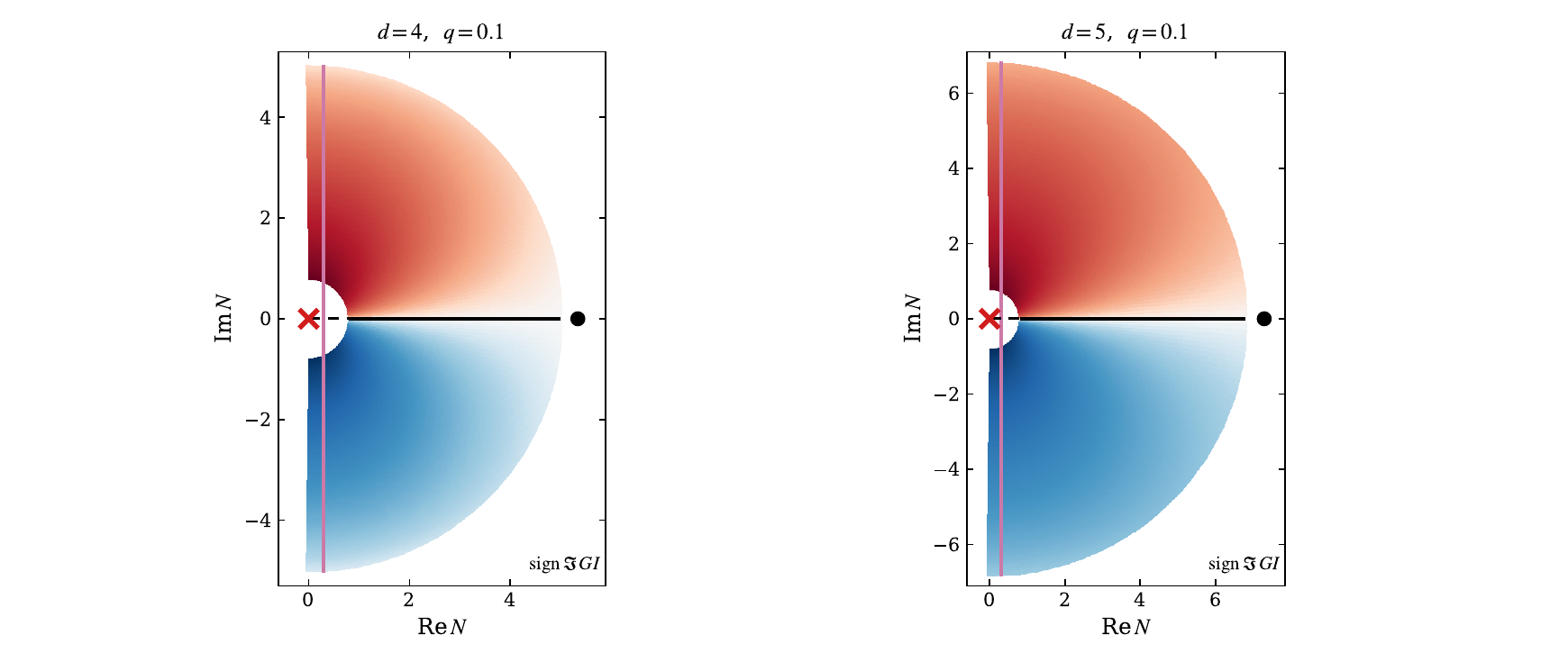}
\caption{Lines of real action $I(N)$ arriving at the origin for 4d/5d axion wormholes in kinetic gauge (black). According to section \ref{subsect2.5.4gen} one should track when those lines reach $H(\mathcal{A}_\kappa)<2\abs{I_0}$. The color reflects the phase.}
\label{fig:zerophase45}
\end{figure}

Figure \ref{fig:zerophase45} shows the ascent lines emerging from the origin for any real saddle. Their degeneracy may be broken by slightly complexifying $G$, however, this will not be needed for our present argument. The higher necklaces have actions $kI_0$ with $k\geq 2$. For some ascent line $\mathcal{A}_\kappa$ to reach the origin (contributing to the defining gravitational path integral) and to have originated from some higher necklace, the ascent line emerging from the origin may not enter any region where $H(\mathcal{A}_\kappa)<2\abs{I_0}$ prior to reaching the saddle (otherwise descent into the saddle is impossible). We now show that any contributing $\mathcal{A}_\kappa$ in fact does enter any region where $H(\mathcal{A}_\kappa)<2\abs{I_0}$, prior to reaching the saddle. 

\begin{figure}[h]
\centering
{\includegraphics[width=\textwidth]{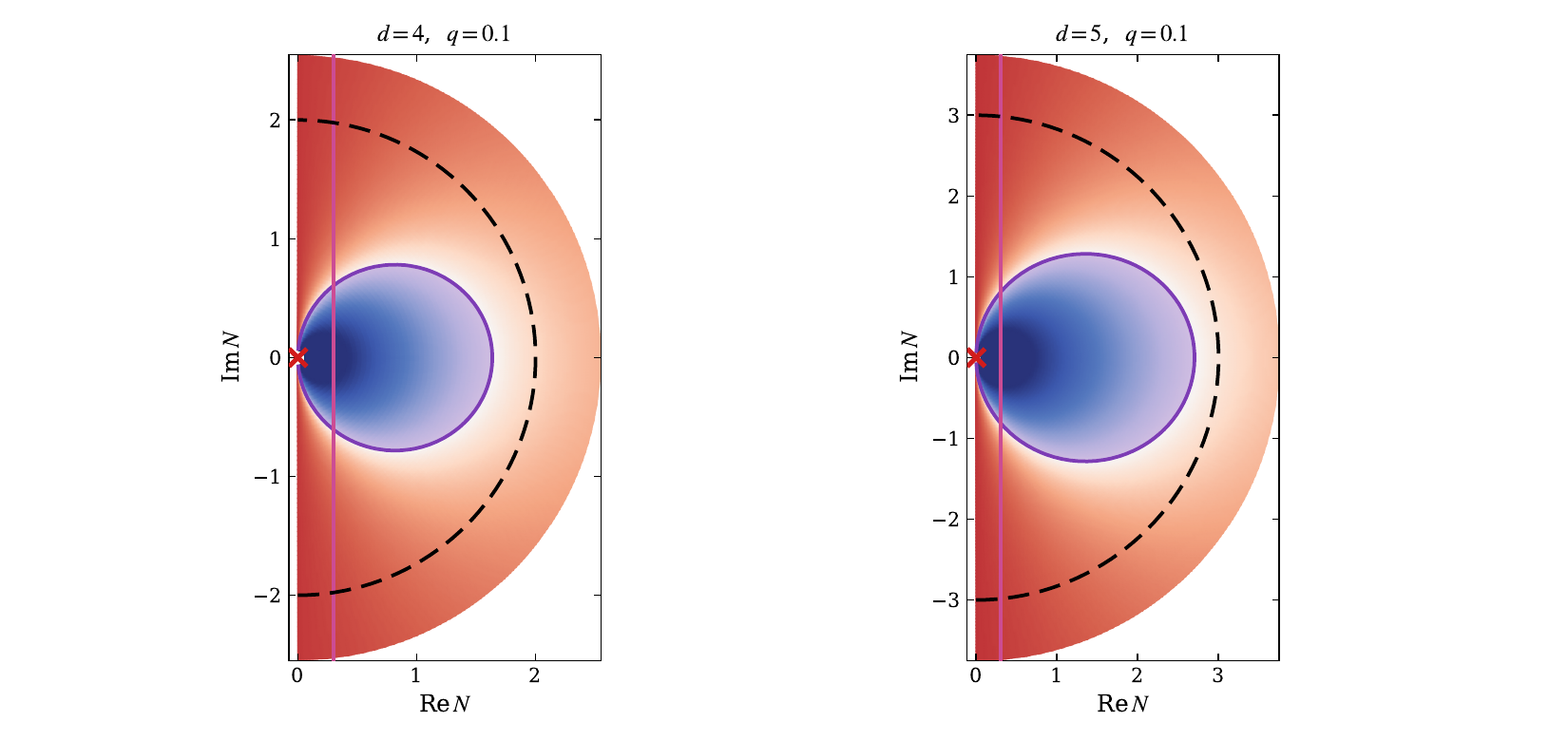}}
\caption{Lines of constant height $H(N)$. Any ascent line ending at $N=0$ (so, contributing to the path integral along the defining contour) necessarily crosses the $H(N)=2\abs{I_0}$ locus (purple). Thus, there are points for every contributing ascent line $\mathcal{A}_\kappa$ where $H(\mathcal{A}_\kappa)<2\abs{I_0}$. This proves that these ascent lines could not originate from $k\geq 2$ necklace solutions. The dashed line (black) is some contour to which the defining contour may be deformed, along which $H(N)<2\abs{I_0}$, fundamentally ruling out contributions from $k\geq 2$ necklaces. The color reflects the height.}
\label{fig:ring}
\end{figure}
Figure \ref{fig:ring} shows numerically for 4d/5d axion wormholes a line with $H(N)=2\abs{I_0}$ (purple) that any contributing $\mathcal{A}_\kappa$ must cross. This rules out contributions from $k\geq 2$ necklaces in the gravitational path integral, thus supporting our claim that the GH saddle dominates, and concluding our discussion.

\section{Magnetic necklaces}\label{sect4:magnetic}
We again consider Euclidean FLRW metric with closed slices
\begin{align}
    \d s^2 =  \d\tau^2 + a(\tau)^2\ \d\Omega_{d-1}^2\,.\label{4.1}
\end{align}
Let us recall the special characteristics of axions that enable them to support a Euclidean cosmological wormhole/necklace \cite{Giddings:1987cg,Hawking:1988ae,Giddings:1988wv,Lavrelashvili:1987jg}. The (first and second) Friedmann equations in Euclidean signature are:\footnote{Here $\theta$ denotes an angle on the $d-1$ sphere $\Omega_{d-1}$ with unit metric component $\gamma_{\theta\theta } = 1$.}
\begin{align}
    \frac{\dot{a}^2}{a^2}  = \frac{1}{a^2} - H^2 +\frac{16\pi G }{(d-1)(d-2)}T_{\tau\tau}\,,\quad \frac{\ddot{a}}{a} = -H^2-\frac{(d-3)}{(d-1)(d-2)}8\pi G\,T_{\tau\tau}+\frac{8\pi G}{(d-2)}\frac{T_{\theta\theta}}{a^2}\,.\label{4.2}
\end{align}
Euclidean wormholes are characterized by the existence of a throat, where $\dot{a}=0$, $\ddot{a}>0$. A sufficiently \textbf{negative energy} $T_{\tau\tau}$ results in such a Euclidean cosmological wormhole. This is of course nonstandard. For instance, a scalar $\phi$ field with an ordinary sign kinetic term and potential $V(\phi)$ has the stress tensor:
\begin{align}
    T_{\tau\tau} = \frac{1}{2}\dot{\phi}^2-V(\phi)\,, \quad \frac{T_{\theta\theta}}{a^2} = -\frac{1}{2}\dot{\phi}^2-V(\phi)\,.
\end{align}
The Friedmann equations \eqref{4.2} now do not allow for configurations with $\dot{a}=0$, $\ddot{a}>0$. Indeed, imposing that $\dot{a}=0$ one finds
\begin{equation}
    \frac{\ddot{a}}{a} = -\frac{1}{a^2}-\frac{8\pi G}{d-2}\dot{\phi}^2<0\,.
\end{equation}
Axions are different: they have an opposite sign kinetic term (when the action is viewed as the dual of the $(d-1)$ form) \cite{Giddings:1987cg}. Similarly, an imaginary massless scalar (such as timelike Liouville theory \cite{Anninos:2024iwf,Harlow:2011ny,Usciati:2025cdn}, depending on the contour) has an opposite sign kinetic term. In this section, we will describe solutions in Yang-Mills theory that also satisfy these criteria.\footnote{We thank Edward Witten for bringing the 4d solution to our attention, from which the generalization to $d>4$ naturally follows.} A nice feature is that their stress energy redshifts as $a^{-4}$ in all dimensions.

\textbf{Yang-Mills solution} The sphere $\Omega_{d-1}$ equals the coset space:
\begin{align}
    S^{d-1} = SO(d)/SO(d-1)\,.
\end{align}
We consider $SO(d-1)$ Yang-Mills on an FLRW background \eqref{4.1}. Identifying the gauge field with a spin connection automatically solves the Yang-Mills equations on Einstein backgrounds \cite{Charap:1977re,Charap:1977ww}. Gauge-field-supported wormholes of this type were constructed in $d=4$ in \cite{Hosoya:1989zn,Gupta:1989bs,Rey:1989th}, though our interpretation will be different. We thus consider the following (purely magnetic) gauge field:\footnote{In $d=4$ this configuration is the meron of \cite{deAlfaro:1976qet}.}
\begin{align}
    A = \frac{1}{2}\omega^{ab}\, t_{ab}\,.
\end{align}
Here, $\omega^{ab}$ is the spin connection on the unit sphere and $t_{ab}$ are the generators of $\mathfrak{so}(d-1)$. By Cartan's structure equation and using the fact that $\Omega_{d-1}$ is maximally symmetric, one computes:
\begin{align}
    \d A=\frac{1}{2}R^{ab}\,t_{a b} - \frac{1}{2}\omega^a_c\wedge \omega^{cb}\,t_{ab}\,,\quad R^{ab} = \frac{1}{2}R^{ab}{}_{cd}\,e^c\wedge e^d = e^a \wedge e^b\,.
\end{align}
Using the $\mathfrak{so}(d-1)$ algebra one furthermore finds:
\begin{align}
     A \wedge A  = \frac{1}{8}\omega^{ab}\wedge \omega^{cd}\, [t_{ab},t_{cd}] = \frac{1}{2}\omega^a_c\wedge  \omega^{cb}\,t_{ab}\,.
\end{align}
Therefore one finds the simple field strength:
\begin{align}
    F = \frac{1}{2}e^a \wedge e^b\, t_{ab}.
\end{align}
The Lorentzian stress-energy tensor associated with this field strength takes the form of a perfect fluid. Restoring $N$ dependence via $\tau\to N\tau$:
\begin{align}
    N^{-2} T_{\tau \tau} = -\frac{(d-1)(d-2)}{4g_\text{YM}^2}\frac{1}{a^4}\,, \quad T_{ii} = \frac{(d-2)(5-d)}{4g_\text{YM}^2}\frac{1}{a^2}\,,\quad T=\frac{(4-d)(d-2)(d-1)}{4g_{YM}^2}\frac{1}{a^4}\,.
\end{align}
In 4d the field is therefore classically conformal ($T=0$), as indeed expected of 4d Yang-Mills theory.

\textbf{Gravitational solution} In minisuperspace, this YM configuration adds a dimension-independent $1/a^2$ contribution to the potential $V(a)$, which for axions was discussed in equation \eqref{eq:ss_high_d}:
\begin{equation}
    I=\frac{\Omega_{d-1}(d-1)(d-2)}{8\pi G}\int_0^1\d\tau\, a^{d-3}\bigg\{-\frac{\dot a^2}{2N}+\frac{N}{2}\bigg(-1+H^2 a^2+\frac{4\pi G}{g_\text{YM}^2}\frac{1}{a^2}\bigg)\bigg\}\,.
\end{equation}
Crucially, the sign of the $1/a^2$ correction is identical to the sign of the axion $1/a^{2(d-2)}$ correction. This creates a wall in the potential around $a\to 0$, and hence results in a bouncing solution. So: a Euclidean necklace. The Friedmann equation (the $N$ equation of motion) indeed admits wormhole solutions: 
\begin{equation}
    \frac{1}{N^2}\frac{\dot{a}^2}{a^2}  = \frac{1}{a^2} - H^2 -\frac{4\pi G}{g_\text{YM}^2}\frac{1}{a^4}\quad \to \quad a^2= \frac{1+ \sqrt{\Delta}\cos(2H N \tau)}{2 H^2}, \quad \Delta = 1- H^2 \frac{16\pi G}{g_\text{YM}^2}\,,\quad 0\leq \Delta\leq1\,.
\end{equation}
These periodic solutions can be repeated any number of times, with classical saddles:
\begin{equation}
    N=\frac{\pi k}{H}\,,\quad 0\leq \tau \leq 1\,.
\end{equation}
The action of one oscillation involves Legendre functions of the first kind, and interpolates between (minus) the dS entropy at $\Delta=1$ and zero at $\Delta=0$:
\begin{align}
    I= \frac{1}{H^{d-4}}\frac{(d-1)\,2^{(5-d)/2}\,\pi^{(d+2)/2}\,}{\Gamma(d/2)}\frac{1}{g_\text{YM}^2}\,(1-\Delta)^{(d-5)/4}\bigg\{P_{\frac{d-5}{2}}\bigg(\frac{1}{\sqrt{1-\Delta}}\bigg) - P_{\frac{d-1}{2}}\bigg(\frac{1}{\sqrt{1-\Delta}}\bigg)\bigg\}\,.
\end{align}

We remark finally that in kinetic gauge $N\to a^{d-3}N$, this YM contribution to $V(a)$ scales as $a^{2d-8}$. In particular in four dimensions the YM contribution becomes simply a constant in units where $H=1$:
\begin{equation}
    \boxed{I_\text{4d} = \frac{3\pi}{2 G}\int_0^1 \d\tau\, \bigg\{-\frac{\dot a^2}{2N} +\frac{N}{2}\bigg(-a^{2}+a^{4} + \frac{4\pi G}{g_\text{YM}^2} \bigg)\bigg\}\,}\label{4.14YMaction}
\end{equation}
This case could be studied analytically. The $a$ equations of motion are solved in terms of Jacobi elliptic tangent functions $\text{sc}(u,m)$, which are indeed (doubly) periodic functions:
\begin{equation}
    \frac{\ddot{a}}{N^2}-a+2a^3=0\,\quad \to \quad a =\sqrt{C^2-1}\,\mathrm{sc} \Big(C N \tau\, \Big|\, 2-\frac{1}{C^2}\Big)\,.\label{4.15sol}
\end{equation}
The special case $g_\text{YM}^2=16\pi G$ results in an Einstein static universe, where $\dot{a}=0$. This corresponds to a particle at rest at the bottom of the quartic potential ($a_\text{rest}=1/\sqrt{2}$). The propagators $\mathcal{G}_N(a_\text{rest}\rvert a_\text{rest})$ and $\mathcal{G}_N(a_\text{rest}\rvert -a_\text{rest})$ are the topic of Coleman's standard example for the study of instantons in QM \cite{Coleman:1978ae}.

One may repeat the numerical analysis of section \ref{sect3.3kineticnum} using the algorithm of section \ref{subsect2.5.4gen} to conclude that the contour $N=\varepsilon+\i \mathbb{R}$ picks up the GH saddle as the leading contribution.

\section{Concluding remarks}\label{sect5:concl}
We emphasized a problem with the Euclidean gravitational path integral, reminiscent of the conformal mode problem, but pertaining to genuine classical solutions. For certain special matter contents (with wrong sign kinetic term) there exist cosmological solutions that oscillate indefinitely in Euclidean space. This means that the Euclidean gravitational path integral is divergent, and therefore cannot reasonably be considered the correct definition of quantum gravity.

Working in minisuperspace, we considered two alternative definitions of quantum gravity which (from this perspective) seem more reasonable. These options are characterized by \textbf{Lorentzian lapse} contours.
\begin{enumerate}
    \item \underline{$N=\varepsilon+\i \mathbb{R}$} The main point of this work was to show that this contour agrees with the Gibbons-Hawking proposal: that the Euclidean gravitational path integral computes the entropy of the dS static patch. This contour similarly predicts a finite order $1/G$ entropy for an observer in a static patch with negative-kinetic-energy matter (like the axion and YM configurations we have studied). The dominant saddle lives in the $a\to -a$ sector:
    \begin{equation}
    \begin{tikzpicture}[baseline={([yshift=-.5ex]current bounding box.center)}, scale=0.7]
    \pgftext{\includegraphics[scale=1]{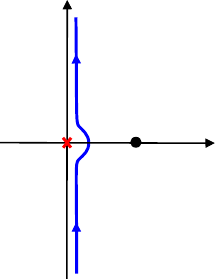}} at (0,0);
    \draw (-1.5,-0.5) node {\color{red}pole};
    \draw (2.5,-0.5) node {dominant saddle};
    \draw (2.5,-1.2) node {GH entropy};
    \draw (1,1.8) node {\color{blue}$N=\varepsilon+\i\mathbb{R}$};
    \draw (-2.9,2) node {$I_{a\to-a}(N)$};
    \end{tikzpicture}
\end{equation}
    \item \underline{$N=-\varepsilon+\i \mathbb{R}$ (or $N = \i\mathbb{R}^+$)} This contour predicts an entropy which is of order one (in $G$) for an observer in the dS static patch. This would mean that the area of the cosmological horizon is not measuring entropy. Even though this seems a radical idea, recent work questions to what degree cosmological area is entropy survives when finite $G$ effects are taken into account \cite{Chen:2026boh,Milekhin:2026tbi,Cui:2026bcd,Harlow:2026pwe}.
    The reason that the contour $N=-\varepsilon+\i \mathbb{R}$ gives no $1/G$ term in the entropy is as follows. Thinking about the trace of the density matrix as integrating over $a_0=a_1$, for $a_0<a_{\text{max}}$ there are two Euclidean saddle points in the $N$ plane close to $N=0$, of positive and negative values of $N$, corresponding to $k=0$. They differ by the orientation of time.  When $a_0$ approaches $a_{\text{max}}$ the proper time of the corresponding saddle goes to zero. This contour picks the saddle with $N<0$. The result of integration over $a_0$ gives in terms of $1/G$ dependence a vanishing entropy, since it is dominated by $a_0\to a_\text{max}$. For a contour $N = \i\mathbb{R}^+$, there is a similar conclusion, but its origin is in the $N=0$ endpoint of the integration contour, rather than a saddle point.
    \begin{equation}
    \begin{tikzpicture}[baseline={([yshift=-.5ex]current bounding box.center)}, scale=0.7]
    \pgftext{\includegraphics[scale=1]{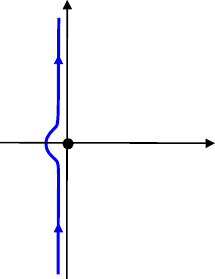}} at (0,0);
    \draw (1.5,-0.5) node {dominant saddle};
    \draw (1.5,-1.2) node {small entropy};
    \draw (1.1,1.8) node {\color{blue}$N=-\varepsilon+\i\mathbb{R}$};
    \draw (-2.9,2) node {$I_{a\to +a}(N)$};
    \end{tikzpicture}
\end{equation}
\end{enumerate}

We end with various comments. Firstly, the minisuperspace approximation is of course not complete. In particular the Picard-Lefschetz analysis may well change beyond minisuperspace. See \cite{Ivo:2026ijv} for an analysis of fluctuations on these spacetimes. Here we simply note that the fact that Lorentzian contours in minisuperspace give reasonable answers is encouraging to undertake an analysis beyond minisuperspace.

Secondly, we have described things from the perspective of a global slicing ADM quantum mechanics. It would of course be interesting to describe this immediately from the perspective of a static-patch-type quantum mechanics, describing an observer's density matrix \cite{Ivo:2024ill} directly. Obviously, such a quantum mechanics is very hard to come by, as its microstates would be a microscopic realization of dS entropy.

Thirdly, it may be interesting to understand if Kontsevich-Segal-Witten in the no-boundary state \cite{Kontsevich:2021dmb,Witten:2021nzp,Hertog:2024nbh,Hertog:2023vot, Lehners:2021mah} has something to say about necklaces, as it did for a chain of spheres $q\to 0$. Our configurations are not allowable according to the KSW criterion: avoiding the big-bang at $a=0$ as in equation \eqref{2.59contour}, one infinitesimally passes through an un-allowable region, crossing for instance $a^2=-\varepsilon^2<0$. However, they may still be physical, as they seem to obey the weaker ``spectral KSW condition'' \cite{Caminiti:2026efx} which is a condition on the well-definedness of one-loop matter partition functions on these complex metrics, allowing for more general contour rotations of the fields.\footnote{We thank Aidan Herderschee for checking the spectrum of a scalar field on this background numerically and in certain limits, providing evidence for this.} 

Relatedly, following \cite{Chen:2020tes,Fumagalli:2024msi}, one could propose the following stability criterion for a general FLRW-type background with scale factor $a(T)$ and spatial manifold $\Sigma$. Imagine we couple conformal matter to this background. By a naive Weyl transformation, one maps to $\Sigma \times I$ with time coordinate $\d\eta = \d T/a(T)$. The Euclidean length of the interval is
\begin{equation}
    \beta = \i \int \frac{\d T}{a(T)}\,.
\end{equation}
For $\text{Re}\,\beta \leq 0$ the CFT partition function diverges, rendering the FLRW spacetime unstable to quantum fluctuations.

How about our contour \eqref{2.59contour}? Naively, $a(T)$ changes sign due to the Lorentzian detour around the singularity, resulting in a troubling $\text{Re}\, \beta=0$.\footnote{We thank Victor Gorbenko for raising this concern.} Take the generator of conformal time transformations $H_\eta=a(T) H_T$. If $\text{Re}\,a(T)>0$ the generator of $\eta$ transformations remains bounded from below, which keeps the stability condition sensible. When $\text{Re}\,a(T)<0$, negative $\eta$ flow arises. However, we notice that simultaneously $H_\eta$ becomes unbounded from above, such that negative $\eta$ evolution actually is the stable case. We propose to think about \eqref{2.59contour} as the product of the contour before and after winding around the singularity. Each part by itself is a bounded operator, and hence so is the product. The alternative would be to conclude that a configuration with $a(T)<0$ everywhere, which is gauge-equivalent to one with $a(T)>0$ everywhere (these describe the same metric), would be an unstable background.\footnote{The GH saddle, in our current interpretation as an $a_\text{max}\to -a_\text{max}$ solution, would also not prepare the global vacuum state $\beta=\infty$ but rather the maximally mixed state $\beta=0$. This would be an extremely troubling conclusion.} Note that the $\pi$ contour \cite{Chen:2020tes,Fumagalli:2024msi} would be to go from $\tau=0$ to $1$ in a straight line in \eqref{2.59contour}. Along this contour $a(\tau)>0$. 

Finally, consider the universal area law scaling for the 3d axion wormhole action \eqref{univ}, see section \ref{sect:3.3}.
It would be interesting to understand if a version of this argument generalized to higher dimensions. Is there any contour in the complex $a$ plane for which the on-shell action (see section \ref{sect3:5daxion}) exactly matches the $q=0$ GH action, which deforms to a late-time dS contour? Converse, suppose we take a late-time dS contour and contour deform ``as much as possible'' to the purely Euclidean contour, what wormhole geometry does this produce? We leave this for future investigations.

\section*{Acknowledgments}
We are happy to thank Jan de Boer, Victor Gorbenko, Aidan Herderschee, Victor Ivo, Oliver Janssen, Juan Maldacena, Don Marolf, Henry Maxfield, Nathan Seiberg, and Edward Witten for discussions. We especially thank Sergio Ernesto Aguilar Gutierrez for pointing us to the work of Halliwell and Myers \cite{halliwell1989multiple}. AB was supported by the US DOE and the Leinweber Foundation. JKF is supported by the Marvin L. Goldberger Member Fund at the Institute for Advanced Study and the National Science Foundation under Grant No. PHY-2514611.
VN is supported by DARPA AIQ grant (HR001124S0029).
EYU was supported by the J. Robert Oppenheimer Endowed Fund and the Fund for Natural Sciences. We acknowledge discussions in the workshop ``Observers, wormholes and complex saddles in cosmology'' at the Bernoulli center, at EPFL, Lausanne, CH. The authors used the assistance of Claude and of ChatGPT for numerical simulations, proofreading, and development of arguments. All derivations were carried out and written up by the authors, who take full responsibility. 

\appendix

\section{Five dimensional axion necklaces using proper-time gauge}\label{app:5dproper}
In this appendix, we will study 5d axion wormholes in proper-time gauge (introduced in section \ref{sect3.1:5fproper}). In section \ref{appa1} we present analytically the case of pure gravity (solutions and on-shell actions). In section \ref{app:contours} we show that the Lorentzian lapse contour $N=\varepsilon+\i\mathbb{R}$ picks up the GH sphere saddle as dominating configuration. In section \ref{app:numerics} we study numerically $q>0$ and show that this picture persists, reducing to our analytical analysis for $q\to 0$.

\subsection{Proper-time gauge}\label{sect3.1:5fproper}
Beginning with the action \eqref{eq:axion_action}, we consider FLRW configurations in the standard \textbf{proper-time gauge}:
\begin{equation}
    \d s^2 = N^2 \d \tau^2 + a(\tau)^2 \d \Omega_{d-1}^2\,,\quad 0\leq \tau\leq 1\,.\label{3.2proper}
\end{equation}
Using \eqref{eq:axion_conv}, one finds the following action:
\begin{equation}\label{eq:ss_high_d}
    I = \frac{\Omega_{d-1}(d-1)(d-2)}{8\pi G}\int_0^1 \d\tau\, a^{d-3}\bigg\{-\frac{\dot a^2}{2N} +\frac{N}{2}\bigg(-1+H^2 a^2 + \frac{1}{d-1}\left(\frac{d-2}{d-1}\right)^{d-2} \frac{q^2}{(H a)^{2(d-2)}} \bigg)\bigg\}.
\end{equation}
Compared to the kinetic gauge analysis of section \ref{sect3:5daxion}, we find two differences. First, the kinetic term for the scale factor is non-standard, which makes the quantum mechanical analysis less standard. Secondly, $a\mapsto -a$ is a symmetry of the action only for odd $d$. 
For this reason, we shall consider only $5d$ here. In units $H=1$, the 5d action becomes:
\begin{equation}
    I_\text{5d} = \frac{4\pi}{G}\int_0^1 \d \tau \bigg\{-\frac{a^2\dot a^2}{2N} +\frac{N}{2}\bigg(-a^2+a^4 + \frac{27}{256} \frac{q^2}{a^4} \bigg)\bigg\}\,.\label{3.5action}
\end{equation}

As the kinetic term is non-standard, it is convenient to think of this system as describing quantum mechanics of a particle with a coordinate:
\begin{equation}
    A=a^2\,.
\end{equation}
The action then becomes
\begin{equation}
    \boxed{I_\text{5d}=\frac{4\pi}{G}\int_0^1\d \tau \bigg\{-\frac{\dot{A}^2}{8 N}+\frac{N}{2}\bigg(-A+A^2+ \frac{27}{256} \frac{q^2}{A^2}\bigg)  \bigg\}\,}\label{3.7action}
\end{equation}
This shares many similarities with the 3d action \eqref{2.8action}. The natural analogs of the $a\to -a$ configurations in 3d are, in proper-time gauge \eqref{3.2proper}, the $A\to -A$ configurations. Following section \ref{subsect2.5.4gen} one finds for $N\to 0$
\begin{equation}
\mathcal{G}_N(A_1\rvert A_2)\, \to\, e^{\frac{\pi}{2 N G}(A_1-A_2)^2}\,.\label{3.8lim}
\end{equation}
This has a very similar property to the relevant 3d $N\to 0$ propagators: $\mathcal{G}_N(A\rvert A)$ has no singularity, and instead a saddle at $N=0$. Furthermore, crucially, $\mathcal{G}_N(A\rvert -A)$ does have a singularity for $N\to 0$. The defining contour thus again deforms to a steepest descent contour through the first saddle (at $N=N_0$). 

In the remainder of this section we study the $q=0$ case analytically, and the finite $q$ case numerically.

\subsection{Pure gravity analytically}\label{appa1}
In this section we take the pure gravity limit $q\rightarrow 0$, which can be solved analytically. Rescaling $\tau \to N\tau$ and introducing
\begin{equation}
    B=A-\frac{1}{2} = a^2-\frac{1}{2}\,,\label{a.3}
\end{equation}
the Euclidean action becomes:
\begin{equation}\label{eq:IE_5d}
    I = \frac{\pi}{G}\int_0^N \d \tau \bigg\{-\frac{1}{2}\dot B^2-\frac{1}{2}+2 B^2  \bigg\}\,.
\end{equation}
This resembles the 3d theory \eqref{2.8action} for $q\to 0$. The equations of motion are
\begin{equation}
    \ddot{B}+4 B=0\,.
\end{equation}
The classical dS solution reads
\begin{equation}
    B=\frac{1}{2}\cos(2\tau)\quad \to \quad a=\cos(\tau)\,,\quad N=\pi\,.
\end{equation}
We distinguish two interesting solution sectors.
\begin{enumerate}
    \item \textbf{Standard solutions} Imposing $a(0)=a_1>0\,,a(N)=a_2>0$ one finds
    \begin{equation}
    a^2_+(\tau) = \frac{1}{2}+\left(a^2_1-\frac{1}{2}\right)\frac{\sin(2(N-\tau))}{\sin(2N)}+\left(a^{2}_2-\frac{1}{2}\right)\frac{\sin(2\tau)}{\sin(2N)}\,.
\end{equation}
    In Lorentzian signature, one takes $\tau = \i T$ and $N = \i N_\text{L}$. The Lorentzian solution satisfies $a^2>0$ for every $N_\text{L}$, and so corresponds to a valid real solution for $a(t)$. The Euclidean solution satisfies $a^2(\tau)>0$ for $0\le \tau \le N$ only when $-\pi/2\le N\le \pi/2$. For generic $N$, $a(\tau)$ is complex. Our perspective is that to study $I_{a_0\to a_0}(N)$ one takes real $a(T)$ on the Lorentzian axis, and analytically continues those in $N$. The action becomes
    \begin{equation}
        \frac{G}{2\pi}I_{a_1\to a_2}(N)=-\frac{N}{4}-
    \frac{1}{2}(B_1^2+B_2^2)\,\cot(2N)
    +\frac{B_1 B_2}{\sin(2N)}\,,\quad B_i=a_i^2-\frac{1}{2}\,.\label{a.12bis}
    \end{equation}
    As a check, for $N\to 0$ we find $I_{a_1\to a_2}(N\to 0)\to -\pi (B_1-B_2)^2/(2 G N)$, consistent with formula \eqref{3.8lim}. For computing the GH entropy, we are interested in $a_1=a_2=a_\text{max}=1$. The action now reads:
    \begin{equation}
        \boxed{\frac{2 G}{\pi} I_{a_\text{max}\to a_\text{max}}(N)=-N+\tan(N)\,}\label{a.9}
    \end{equation}
    This is identical to the 3d action for pure dS \eqref{2.37I} modulo $N\to N/2$. Indeed, rescaling $N\to N/2$ and $B\to B/2$ brings equations \eqref{eq:IE_5d} and \eqref{2.8action} in identical form.
    \item \textbf{Bouncing solutions} Based on our analysis of section \ref{sect2:3daxion}, we would like to mimic a solution which goes through the origin in the $a$ plane: $a_0\to 0\to -a_0$. Notice that this involves only $B\geq -1/2$. Indeed, equation \eqref{a.3} suggests that we should view the $B$ quantum theory as having a brick wall at $B=-1/2$:
    \begin{equation}
        \begin{tikzpicture}[baseline={([yshift=-.5ex]current bounding box.center)}, scale=0.7]
    \pgftext{\includegraphics[scale=1]{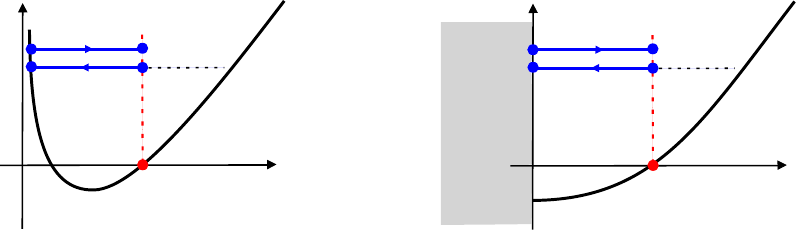}} at (0,0);
    \draw (3,1.5) node {\color{blue}$2$};
    \draw (3,0.4) node {\color{blue}$1$};
    \draw (5,-1.3) node {\color{red}$B_\text{max}$};
    \draw (-1.1,1.65) node {$V(a)$};
    \draw (2.1,-2.4) node {$-\frac{1}{2}$};
    \draw (6.4,-1.3) node {$B$};
    \draw (-0.7,0) node {$\overset{q\to 0}{\longrightarrow}$};
    \draw (1.5,0) node {wall};
    \end{tikzpicture} \label{a.8}
    \end{equation}
    The solution $a_0\to 0\to -a_0$ is translated in $B$ coordinates as a particle that bounces off the wall at $B=-1/2$. In quantum mechanics, we compute the total propagator by first propagating from $a_0\to 0$ and then from $0\to -a_0$. The total action is
    \begin{equation}
        I_{a_0\to -a_0}(N)=I_{a_0\to 0}(N_1)+I_{0\to-a_0}(N-N_1)\,.
    \end{equation}
    The solution is found upon extremizing over $N_1$. In the Hamiltonian formulation, which involves a momentum $P=\dot{B}/N$ as in \eqref{2.23actionlor}, extremizing over $N_1$ imposes that the $a$ particle has the same energy before and after the bounce: $E_1=E_2$. This imposes that after the bounce the momentum $P$ simply flips sign. The saddle is: $N_1=N/2$. A particle with conserved energy indeed takes the same time to reach the wall as it does to come back. Therefore we find:
    \begin{equation}
        \frac{G}{2\pi} I_{a_0\to-a_0}(N)=
    -\frac{N}{4} 
    +\left(a_0^2-a_0^4-\frac{1}{2}\right)\cot(N)
    - \left(a_0^2-\frac{1}{2}\right)\frac{1}{\sin(N)}\,.\label{a.12}
    \end{equation}
    More specifically:
    \begin{equation}
        \boxed{\frac{2G}{\pi} I_{a_\text{max}\to -a_\text{max}}(N)=-N-2\cot(N/2)\,}\label{a.13}
    \end{equation}
    This action has saddles at $N=\pi+2\pi k$ and singularities at $N=2\pi k$. A contour analysis in the next section \ref{app:contours} shows that the Lorentzian contour $N=\varepsilon+\i \mathbb{R}$ is dominated by the saddle $N=\pi$ with entropy $S=-I$ equal to the area of the cosmological horizon of 5d dS space, as is expected:
    \begin{equation}
        S=\frac{\pi^2}{2 G}\,.
    \end{equation}
    In section \ref{app:numerics} we check numerically that bouncing off a wall as described here is the correct $q\to 0$ limit of the particle bouncing off the finite potential wall at $B\to -1/2$ due to $q\neq 0$ axions. See for instance equation \eqref{3.7action} for the $q\neq 0$ potential wall.
\end{enumerate}

\subsection{Deformations of Lorentzian contours}\label{app:contours}
We next perform the steepest descent analysis starting from the Lorentzian contour $N=\varepsilon+\i \mathbb{R}$ for the actions \eqref{a.12bis}/\eqref{a.12}.

\begin{figure}
\centering
\includegraphics[width=\textwidth]{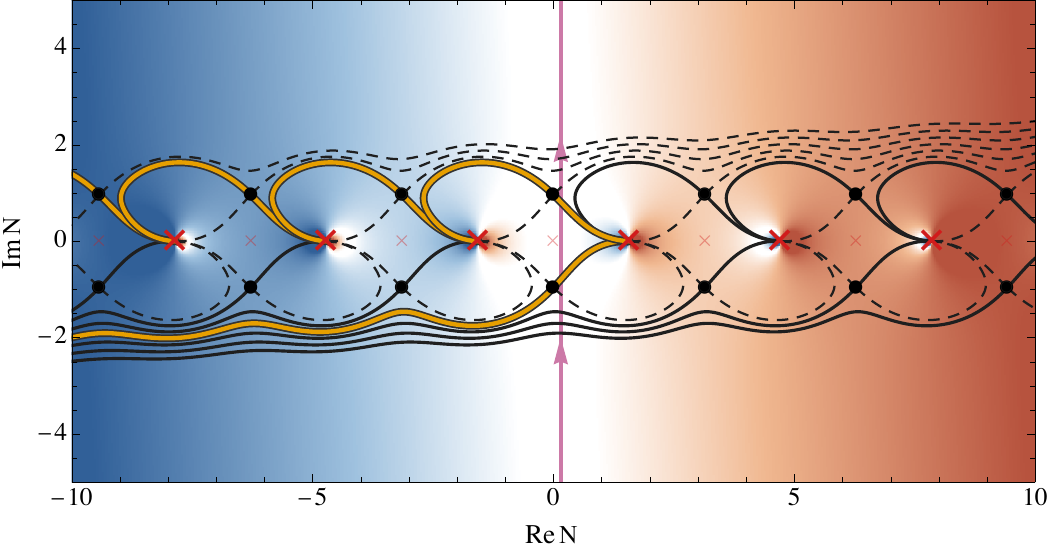}
\caption{Complex $N$ plane for $I_{a_0 \to +a_0}(N)$ for pure dS \eqref{a.13xx}. Poles in the action are red $\cross$'s and saddles are black dots. Steepest descent (ascent) curves are black solid (dashed) lines. The Lorentzian $N=\varepsilon+\i\mathbb{R}$ contour is drawn vertically in magenta and the absolutely convergent contour that it is deformed into is drawn in orange. Here $G$ was slightly complexified to make the ascent/descent lines non-degenerate.}
\label{fig:app1}
\end{figure}

\textbf{Standard solutions} Consider first a trajectory $a_0\to a_0$. The Euclidean action \eqref{a.12bis} matches with the 3d action \eqref{2.37I} (when rescaling the integrating variable $N\to N/2$):
\begin{equation}
    -\frac{2G}{\pi}I_{a_0\to a_0}(N)=N-a_{0\,\text{3d}}^2\tan(N)\,,\quad a_{0\,\text{3d}}=2 a_0^2-1\,.\label{a.13xx}
\end{equation}
The structure of singularities and saddles was discussed in section \ref{sect2.2exact}. This action has singularities at $N=\pi(k+1/2)$. The saddles are:
\begin{equation}
    N=\pm \i\, \text{arccosh}(a_{0\,\text{3d}})+\pi k\,.
\end{equation}
For $a_0>1$, the classical solution has a piece of Lorentzian evolution, hence the imaginary contributions to elapsed time $N$. Expanding around the saddles, one finds that fixed phase lines emerging from these saddles have angles $\pm \pi/4$ with the real $N$ axis. A Lorentzian contour $N= \pm \varepsilon+\i \mathbb{R}$ picks up the saddles with $k\leq 0$. See figure \ref{fig:app1}.
This is a slightly more refined version of the 3d figure \ref{fig:plot_plus_sol_N_plane}. The leading saddle has purely imaginary action, as the $k=0$ solution represents an entirely Lorentzian bra-ket wormhole:
\begin{equation}
    \pm \frac{2G}{\pi}I_{a_0\to a_0}(N)=-\i\,\text{arccosh}(a_{0\,\text{3d}})+\i\,a_{0\,\text{3d}}\sqrt{a_{0\,\text{3d}}^2-1}\,.
\end{equation}

\begin{figure}
\centering
\includegraphics[width=\textwidth]{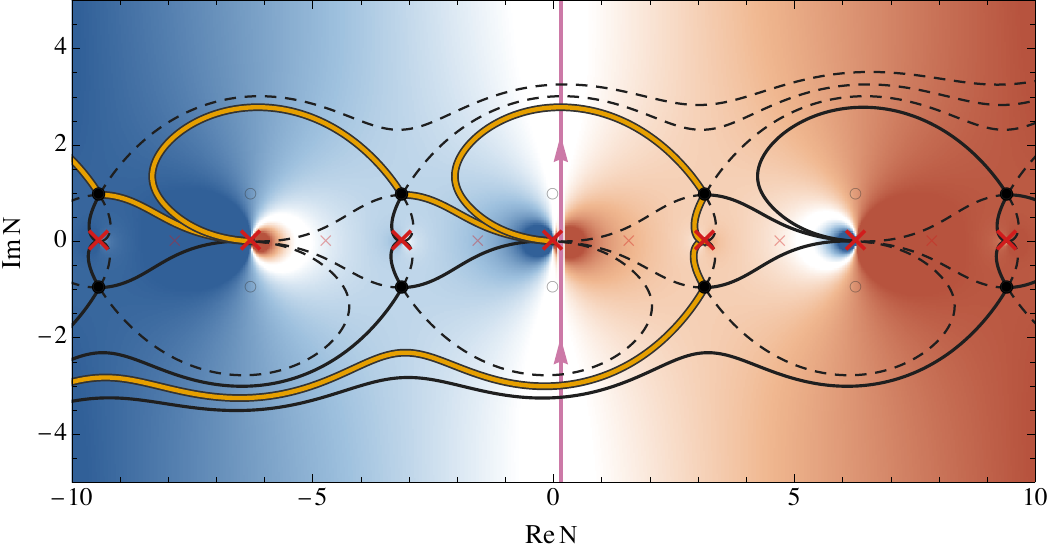}
\caption{Complex $N$ plane for $I_{a_0 \to -a_0}(N)$ for pure dS \eqref{a.17bounce}. Poles in the action are red $\cross$'s and saddles are black dots. Steepest descent (ascent) curves are black solid (dashed) lines. The Lorentzian $N=\varepsilon+\i\mathbb{R}$ contour is drawn vertically in magenta and the absolutely convergent contour that it is deformed into is drawn in orange. Here $G$ was slightly complexified to render the ascent/descent lines non-degenerate. Singularities between the saddles go away for $a_0\to a_\text{mas}$.}
\label{fig:app2}
\end{figure}

\textbf{Bouncing solutions} The action \eqref{a.12} can be rewritten as
\begin{equation}
    -\frac{2G}{\pi} I_{a_0\to-a_0}(N)=N +(a_{0\,\text{3d}}^2+1)\cot(N)+\frac{2 a_{0\,\text{3d}}}{\sin(N)}\,.\label{a.17bounce}
\end{equation}
The singularities are located at $N=\pi k$ and the saddles are located at
\begin{equation}
    N=\pm \i\, \text{arccosh}(a_{0\,\text{3d}})+\pi (2k+1)\,.
\end{equation}
In the case in which we are interested, $a_{0\,\text{3d}}=1$ and the saddles are on the real axis. The only remaining singularities are at $N=2\pi k$. The action is singular at $N=0$, is real for real $N$, and $-I_{a_\text{max}\to -a_\text{max}}(N)$ decreases monotonically to the saddle at $N=\pi$. The steepest ascent contour starting at $N=\pi$ crosses the contour $N=\varepsilon+\i \mathbb{R}$. According to Picard-Lefschetz, this means that the $N=\pi$ saddle contributes to the Lorentzian gravitational path integral. A detailed plot confirms that this is also the dominating saddle. See figure \ref{fig:app2}.

Notice that instead choosing the contour $N=-\varepsilon+\i \mathbb{R}$, the leading contribution is $k=-1$. Hence, the path integral would predict $S=0$ (at order $1/G$). In subsection \ref{app:numerics}, we show numerically that this behavior of the action $I_{a_0\to-a_0}(N)$ persists for finite $q\neq 0$.

\subsection{Numerical analysis at finite q}\label{app:numerics}
\begin{figure}
    \centering
\includegraphics[width=\textwidth]{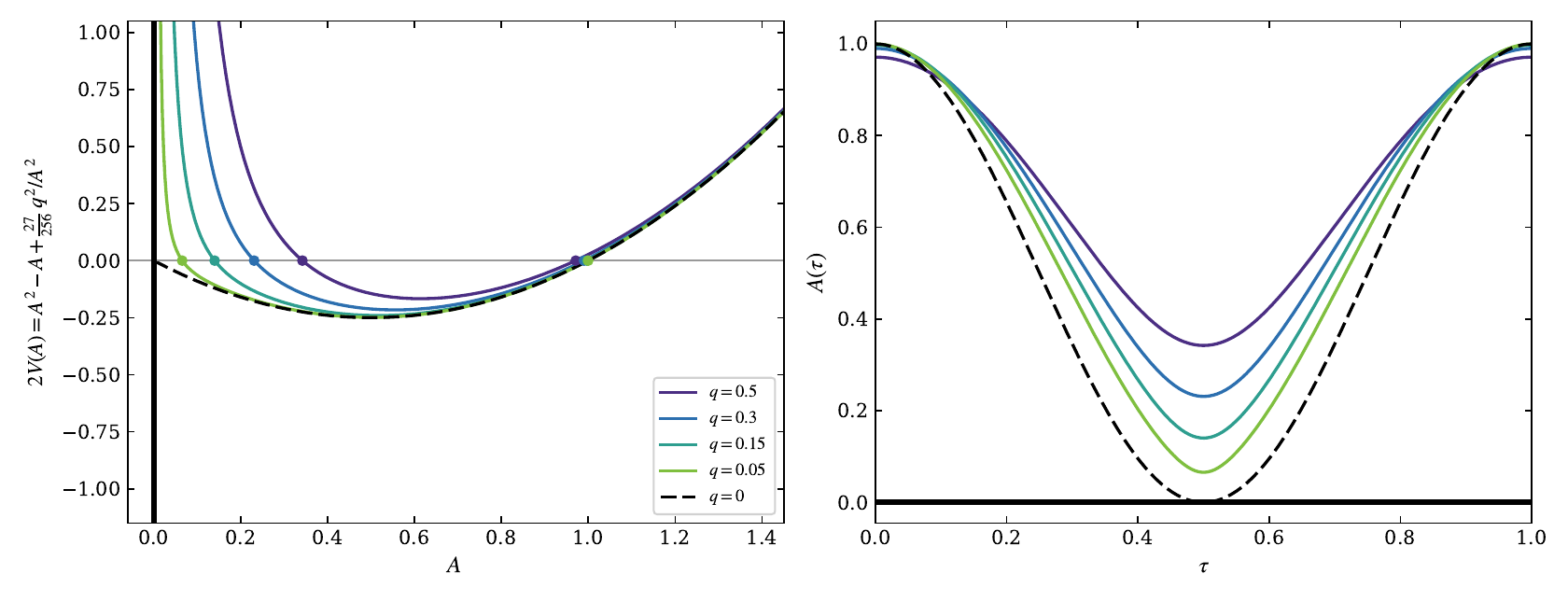}
    \caption{The finite $q$ numerical solution of \eqref{a.18} limits to a solution which bounces off the wall at $A=0$ for $q\to 0$.}
    \label{fig:appnum1}
\end{figure}

Recall equation \eqref{3.7action} (with $A=a^2$ and rescaled $N\to N/2$):
\begin{equation}
    I_\text{5d}=\frac{2\pi}{G}\int_0^1\d \tau \bigg\{-\frac{\dot{A}^2}{2 N}+\frac{N}{2}\bigg(-A+A^2+ \frac{27}{256} \frac{q^2}{A^2}\bigg)  \bigg\}\,.\label{a.18}
\end{equation}
Figure \ref{fig:appnum1} shows the numerical solutions for a particle moving in this potential with boundary conditions $a_\text{max}\to -a_\text{max}$, generalizing the bouncing solution to finite $q$. For $q\to 0$ we found an action \eqref{a.13} for the bouncing solution. The numerical analysis leads to the following conclusions. The finite $q$ solution for $A$ limits to a solution which bounces off a wall at $A=0$ for $q\to 0$, 
confirming the picture \eqref{a.8}. 

\begin{figure}
    \centering
\includegraphics[width=\textwidth]{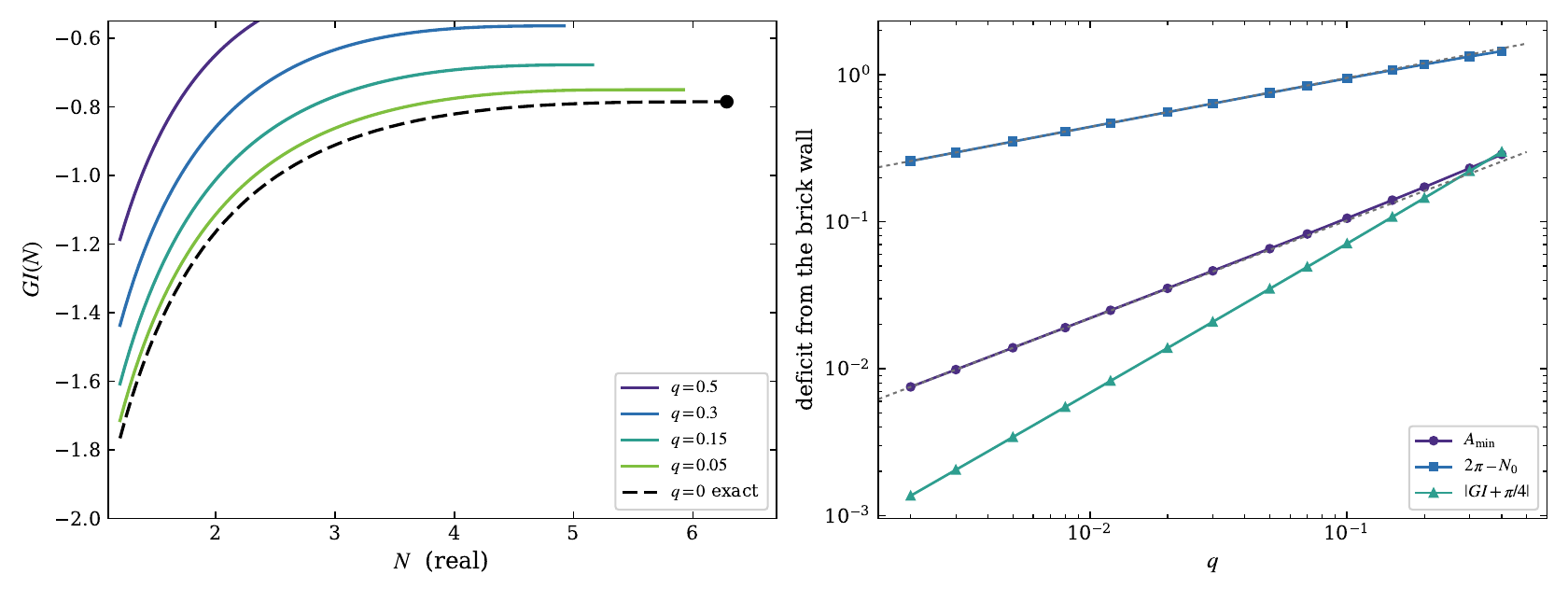}
    \caption{The finite $q$ action \eqref{a.18} approaches the bounce analytic action \eqref{a.13} for $q\to 0$ (left), taking into account rescaling $N\to N/2$. For $q\to 0$ one finds $A_\text{min}\sim q^{2/3}$, $2\pi-N_0\sim q^{1/3}$, and the action deficit scaling with $q$ analytically. Numerics support this (right).}
    \label{fig:appnum2}
\end{figure}

As shown in figure \ref{fig:appnum2}, the numerically evaluated action $I_{a_\text{max}\to -a_\text{max}}(N)$ associated with the bouncing solution matches with the analytic formula \eqref{a.13} when $q\to 0$. Furthermore, the action $I_{a_\text{max}\to a_\text{max}}$ of the standard solution limits to \eqref{a.9}. 

The difference for finite $q$ is the appearance of branch points off the real $N$ axis, identified analytically in 3d below equation \eqref{2.33Ia-a}. These branch points represent values of $N$ where the bouncing and standard solutions coincide, as we see explicitly in 3d in equation \eqref{2.52amin}. The associated branch cuts connect the two sheets of the solution/action, namely $a_\text{max}\to  a_\text{max}$ and $a_\text{max}\to -a_\text{max}$. In 3d we have at times called these the $a_+$ and $a_-$ solutions. The general situation, for general potential, general gauge, or general dimension, as discussed in section \ref{sect3.3kineticnum}, is a multi-sheeted solution space, on which one must perform a Picard-Lefschetz analysis.

\bibliographystyle{ourbst.bst}
\bibliography{main}

\end{document}